\documentclass[preprintnumbers,aps,prd,showpacs,nofootinbib,notitlepage]{revtex4-1}
\usepackage{aas_macros}
\usepackage{amsmath,amssymb,amsfonts}
\usepackage[dvipdfmx]{graphicx}
\usepackage{enumerate} 
\usepackage{color} 
\usepackage{bm}

\newcommand{\bfv}{\mbox{\boldmath$v$}}

\newcommand{\bfx}{\mbox{\boldmath$x$}}
\newcommand{\bfk}{\mbox{\boldmath$k$}}
\newcommand{\bfp}{\mbox{\boldmath$p$}}
\newcommand{\bfq}{\mbox{\boldmath$q$}}

\newcommand{\bfu}{\mbox{\boldmath$u$}}

\newcommand{\bfzero}{\mbox{\boldmath$0$}}

\newcommand{\PkCross}{P_{\rm cross}}
\newcommand{\PkSPT}{P_{\rm\scriptscriptstyle SPT}}

\newcommand{\rhom}{\rho_{\rm m}}

\newcommand{\Prob}{\mbox{Prob}}

\definecolor{RedWine}{rgb}{0.743,0,0}

\begin{document}
\title{GridSPT: Grid-based calculation for perturbation theory of large-scale structure}

\author{Atsushi Taruya}
\affiliation{Center for Gravitational Physics, Yukawa Institute for Theoretical Physics, Kyoto University, Kyoto 606-8502, Japan}
\affiliation{Kavli Institute for the Physics and Mathematics of the Universe (WPI), The University of Tokyo Institutes for Advanced Study, The University of Tokyo, 5-1-5 Kashiwanoha, Kashiwa, Chiba 277-8583, Japan}

\author{Takahiro Nishimichi}
\affiliation{Kavli Institute for the Physics and Mathematics of the Universe (WPI), The University of Tokyo Institutes for Advanced Study, The University of Tokyo, 5-1-5 Kashiwanoha, Kashiwa, Chiba 277-8583, Japan}

\author{Donghui Jeong}
\affiliation{Department of Astronomy and Astrophysics and Institute for Gravitation and the Cosmos, The Pennsylvania State University, University Park, PA 16802, USA}
%
\date{\today}
\begin{abstract}
Perturbation theory (PT) calculation of large-scale structure has been used to interpret the observed non-linear statistics of large-scale structure at the quasi-linear regime. In particular, the so-called standard perturbation theory (SPT) provides a basis for the analytical computation of the higher-order quantities of large-scale structure. Here, we present a novel, grid-based algorithm for the SPT calculation, hence named GridSPT, to generate the higher-order density and velocity fields from a given linear power spectrum. Taking advantage of the Fast Fourier Transform, the GridSPT quickly generates the nonlinear density fields at each order, from which we calculate the statistical quantities such as non-linear power spectrum and bispectrum. Comparing the density fields (to fifth order) from GridSPT with those from the full N-body simulations in the configuration space, we find that GridSPT accurately reproduces the N-body result on large scales. The agreement worsens with smaller smoothing radius, particularly for the under-dense regions where we find that 2LPT (second-order Lagrangian perturbation theory) algorithm performs well.
\end{abstract}

\preprint{YITP-18-73}
\maketitle

\section{Introduction}
\label{sec:intro}

Large-scale matter inhomogeneities in the Universe, as partly traced by the spatial distribution of galaxies and clusters, are thought to have evolved from tiny fluctuations under the influence of gravity and cosmic expansion. The statistical properties of large-scale structure therefore contains rich cosmological information. This is why the large-scale structure observations has been playing a major role to improve our understanding of the cosmology, and there are ongoing/upcoming observations aiming at precisely measuring the power spectrum and the two-point correlation function of large-scale structure over a gigantic cosmological volume. After an idealistic observation of cosmic microwave background anisotropies by Planck \cite{Planck2015_XIII}, large-scale structure would be the best cosmological probe especially to pin down the late-time evolution of the Universe. 

In confronting with future precision observations, an accurate modeling of the large-scale structure is crucial for the unbiased estimation of the cosmological parameters. While the cosmological $N$-body simulation is an essential and standard tool to deal with dark matter and halo clustering even at small scales, the analytic treatment with perturbation theory (PT) is powerful and indispensable to the statistical prediction at quasi-linear scales \cite{Bernardeau:2001qr}. It is at the quasi-linear scales where the measurements of baryon acoustic oscillations and redshift-space distortions, which probe the nature of cosmic acceleration as well as gravity on cosmological scales, are undertaken with large-volume galaxy redshift surveys. This is why numerous works on PT treatment have been done (e.g., \cite{Jeong:2006xd,Crocce:2007dt,Bernardeau:2008fa,Taruya:2007xy,Taruya:2010mx,Matsubara:2007wj,Pietroni:2008jx}). One advantage of the PT calculation is that it provides a way to directly compute the statistical quantities. This is indeed the standard PT (SPT) treatment, in which the higher-order PT kernels of the density and velocity fields (the building blocks of PT calculations defined in the Fourier space) are analytically constructed with recursion relations (e.g., \cite{Bernardeau:2001qr,Goroff:1986ep}), and are used to give analytical expressions for the higher-order corrections to the power spectrum or bispectrum (but see for Refs.~\cite{Taruya2016,Bose_Koyama2016} for the cases when the analytical construction of PT kernels is intractable). Although the SPT expansion is known to have a poor convergence in predicting statistical quantities (e.g., \cite{Crocce:2005xy,Taruya:2009ir}), the framework of the SPT treatment is still useful and in fact used in the re-summed PT scheme which improves the convergence of PT expansion \cite{Crocce:2012fa,Taruya:2012ut}. 

So far, most of the works with the SPT takes advantage of its analytical basis, and has focused on the direct calculation of the statistical quantities (but see Ref.~\cite{roth/porciani:2011,Tassev2014}) such as nonlinear matter power spectrum and bispectrum. In this paper, we present a novel GridSPT method to generate the order-by-order nonlinear density fields from a realization of the Gaussian linear density field. Because the Gaussian random field is used to generate the initial conditions for $N$-body simulations, this method makes possible a face-to-face comparison between SPT and the $N$-body simulation. We study in detail the properties of the generated density fields from both statistical and morphological point-of-view.

This paper is organized as follows. We begin by briefly reviewing the basis of SPT treatment in Sec.~\ref{sec:SPT}. We then present a Fast-Fourier-Transform based method to compute the higher-order density fields from the SPT on grids starting with random density field (Sec.~\ref{sec:gridSPT}). The results of the grid-based PT calculations up to the fifth order are shown in Sec.~\ref{sec:demonstration}, and the structure and statistics of the density fields are compared with $N$-body simulations in detail. In particular, we study the cross-correlation properties of the grid-based PT and $N$-body density fields in both real and Fourier space, and discuss the pros and cons of the SPT prediction. Finally, Sec.~\ref{sec:conclusion} is devoted to conclusion and discussion on the possible application of grid-based PT calculations.

\section{Standard perturbation theory}
\label{sec:SPT}

\subsection{Basic equations}
\label{sec:SPT_basics}

The main objective of this study is to find an accurate model for the large-scale matter inhomogeneities in the cold dark matter (CDM) dominated Universe. Rigorously speaking, the gravitational evolution of such a system is described by the Vlasov-Poisson equation in a cosmological background. Starting with a cold initial condition, the motion of dark matter distribution basically follows a single-stream flow at an early phase of structure formation, and, in such a regime, the system is described by the pressureless fluid equations coupled with the Poisson equation. Then the basic equations for mass density field become (e.g., \cite{Bernardeau:2001qr})
\begin{align}
&\frac{\partial \delta}{\partial t} +
\frac{1}{a}\nabla\cdot[(1+\delta)\bfv]=0,
\label{eq:eq_continuity}\\
&\frac{\partial \bfv}{\partial t} + H\,\bfv+
\frac{1}{a}(\bfv\cdot\nabla)\cdot\bfv=-\frac{1}{a}\nabla\psi,
\label{eq:eq_Euler}
\\
&\frac{1}{a^2}\nabla^2\psi=4\pi\,G\,\rhom\,\delta.
\label{eq:eq_Poisson}
\end{align}
Note that the single-stream flow approximation is eventually violated in the nonlinear regime where the multi-stream flow must be included. Although this can basically happen at small scales, and hence the single-stream approximation is expected to give an accurate description of the non-linear mode coupling in the weakly non-linear regime, recent studies \cite{Bernardeau:2012ux,Blas:2013aba,Nishimichi:2014rra} have advocated that the small-scale multi-stream flows can give an impact on the prediction at large scales, and their effects need to be accounted for (see \cite{Taruya_Colombi2017,McDonald_Vlah2018} for an attempt in 1D). We shall later discuss the quantitative impact of the breakdown of the single-stream flow on the prediction of the large-scale matter distribution.

Adopting Eqs.~(\ref{eq:eq_continuity})-(\ref{eq:eq_Poisson}) as our basic equations, we further impose the irrotationality of the mass flow consistent with the standard picture of structure formation. Then the velocity field follows the potential flow, and is characterized by the scalar quantity. We define $\theta\equiv-\nabla\cdot\bfv/(faH)$ with $f$ being the linear growth rate defined by $f\equiv d\ln D_+/d\ln a$ with $D_+$ being the linear growth factor.
Further, we introduce a new time variable $\eta\equiv\ln D_+(t)$  
and rewrite Eqs.~(\ref{eq:eq_continuity})-(\ref{eq:eq_Poisson}) with evolution equations for $\delta$ and $\theta$. We obtain
\begin{align}
\frac{d}{d\eta}\left(
\begin{array}{c}
\delta(\bfx)
\\
\\
\theta(\bfx)
\end{array}
\right)+\Omega_{ab}(\eta)\,\left(
\begin{array}{c}
\delta(\bfx)
\\
\\
\theta(\bfx)
\end{array}
\right)=\left(
\begin{array}{c}
{\displaystyle (\nabla\delta)\cdot\bfu+\delta\,\,\theta}
\\
\\
{\displaystyle (\partial_ju_k)(\partial_ku_j)+(\nabla\theta)\cdot\bfu}
\end{array}
\right),
\label{eq:basic_PT_eqs}
\end{align}
where the quantity $\bfu$ is the {\it reduced} velocity field defined by 
$\bfu\equiv-\bfv/(f\,aH)$. Under the irrotational flow assumption, it is related to the velocity divergence $\theta$ through
\begin{align}
\bfu(\bfx)=
\nabla\left[\nabla^{-2}\theta(\bfx) \right]
=\int\frac{d^3\bfk}{(2\pi)^3}\left(-\frac{i\,\bfk}{k^2}\right)\,\theta(\bfk)\,e^{i\bfk\cdot\bfx}.
\label{eq:u_and_theta}
\end{align}
In Eq.~(\ref{eq:basic_PT_eqs}), the quantity $\Omega_{ab}$ is a time-dependent $(2\times2)$ matrix: 
\begin{align}
\Omega_{ab}(\eta)=\left(
\begin{array}{cc} 
0 & -1 
\\
\\
{\displaystyle -\frac{4\pi\,G\,\rho_{\rm m}}{f^2\,H^2}} &\quad
{\displaystyle \frac{1}{f}\left(2+\frac{\dot{H}}{H^2}+\frac{df}{d\eta}\right) }
\end{array}
\right).
\label{eq:Omga_ab}
\end{align}
Note that in the Einstein-de Sitter (flat, matter-dominated) Universe, this matrix is reduced to 
\begin{align}
\Omega_{ab}(\eta)
\quad \longrightarrow \quad
\Omega_{ab}^{\rm EdS}=\left(
\begin{array}{cc}
0 & \qquad -1
\\
\\
{\displaystyle -\frac{3}{2}} & \qquad {\displaystyle \frac{1}{2}}
\end{array}
\right).
\label{eq:Omega_ab_EdS}
\end{align}

\subsection{Real-space formalism}
\label{sec:real-space_PT}

Eq.~(\ref{eq:basic_PT_eqs}) with (\ref{eq:Omga_ab}) is the basis to develop a systematic perturbative expansion. Since we are interested in the late-time evolution of matter fluctuations, we may consider the perturbations dominated by the linear growing-mode solution. Further, we shall adopt the so-called Einstein-de Sitter approximation, by which the matrix $\Omega_{ab}$ in Eq.~(\ref{eq:basic_PT_eqs}) is replaced with the one in the Einstein-de Sitter Universe, Eq.~(\ref{eq:Omega_ab_EdS}). The PT calculation with Einstein-de Sitter approximation is shown to give a sufficiently accurate prediction in the cosmological model close to the $\Lambda$CDM model (e.g., \cite{Pietroni:2008jx,Hiramatsu_Taruya2009}), and it ensures that the perturbative quantities $\delta$ and $\theta$ can be expanded by a power series of growth factor $D_+=e^{\eta}$. We thus have
\begin{align}
\delta(\bfx) = \sum_n\,e^{n\,\eta}\,\delta_n(\bfx), \qquad
\theta(\bfx) = \sum_n\,e^{n\,\eta}\,\theta_n(\bfx), \qquad
\bfu(\bfx) = \sum_n\,e^{n\,\eta}\,\bfu_n(\bfx). 
\end{align}
Substituting these into Eq.~(\ref{eq:basic_PT_eqs}), the order-by-order calculation leads to (for $n\geq2$)
\begin{align}
\Bigl( \,n\,\delta_{ab}+\Omega_{ab}^{\rm EdS} \Bigr)\,
\left(
\begin{array}{c}
\delta_n(\bfx)
\\
\\
\theta_n(\bfx)
\end{array}
\right) = \sum_{m=1}^n \left(
\begin{array}{c}
{\displaystyle (\nabla\delta_m)\cdot\bfu_{n-m} + \delta_m\,\,\theta_{n-m}}
\\
\\
{\displaystyle [\partial_j(\bfu_{m})_k][\partial_k(\bfu_{n-m})_j]+\bfu_m\cdot(\nabla\theta_{n-m}) }
\end{array}
\right). 
\label{eq:PTexpansion}
\end{align}
This gives the recursion relation for perturbative quantities $\delta_n$ and $\theta_n$: 
\begin{align}
\left(
\begin{array}{c}
{\displaystyle \delta_n(\bfx) }
\\
\\
{\displaystyle \theta_n(\bfx) }
\end{array}
\right) = \frac{2}{(2n+3)(n-1)}\,\left(
\begin{array}{cc}
{\displaystyle n+\frac{1}{2}} & \qquad 1
\\
\\
{\displaystyle \frac{3}{2} } & \qquad n
\end{array}
\right) \sum_{m=1}^{n-1} \left(
\begin{array}{c}
{\displaystyle (\nabla\delta_m)\cdot\bfu_{n-m} + \delta_m\,\,\theta_{n-m}}
\\
\\
{\displaystyle [\partial_j(\bfu_{m})_k][\partial_k(\bfu_{n-m})_j]+\bfu_m\cdot(\nabla\theta_{n-m}) }
\end{array}
\right),
\label{eq:recursion_formula}
\end{align}
for $n\geq2$. For the linear-order quantities ($n=1$), the growing-mode initial condition implies
\begin{align}
\left(
\begin{array}{c}
\delta_1(\bfx)
\\
\\
\theta_1(\bfx)
\end{array}
\right) = \left(
\begin{array}{c}
1
\\
\\
1
\end{array}
\right) \delta_0(\bfx),
\label{eq:recursion_n=1}
\end{align}
where $\delta_0(\bfx)$ is the initial density field.

The real-space recursion relation given above contains the gradient and vector fields in the nonlinear source terms. To analytically evaluate these nonlinear terms, a standard way is to go to Fourier space, and obtain the local expression for the Fourier kernels of perturbations. This is what has been done in the statistical predictions of large-scale structure. That is, we express the Fourier transform of the density and velocity-divergence fields, $\delta_n(\bfk)$ and $\theta_n(\bfk)$, as
\begin{align}
& 
\delta_n(\bfk) =\int \frac{d^3\bfk_1\cdots d^3\bfk_n}{(2\pi)^{3(n-1)}} \delta_{\rm D}(\bfk-\bfk_{1\cdots n})\,F_n(\bfk_1,\cdots,\bfk_n)\,\delta_0(\bfk_1)\cdots\delta_0(\bfk_n),
\label{eq:delta_n_Fourier}
\\
& 
\theta_n(\bfk) =\int \frac{d^3\bfk_1\cdots d^3\bfk_n}{(2\pi)^{3(n-1)}} \delta_{\rm D}(\bfk-\bfk_{1\cdots n})\,G_n(\bfk_1,\cdots,\bfk_n)\,\delta_0(\bfk_1)\cdots\delta_0(\bfk_n),
\label{eq:theta_n_Fourier}
\end{align}
with $\bfk_{1\cdots n}\equiv\bfk_1+\cdots+\bfk_n$. Here, the field $\delta_0$ is the initial linear density field, and the functions $F_n$ and $G_n$ are the Fourier-space PT kernels. Substituting these expressions into Eq.~(\ref{eq:recursion_formula}), one obtains the Fourier-space recursion relation for PT kernels. Writing $\mathcal{F}_a^{(n)}\equiv(F_n,G_n)$, we have
\begin{align}
 \mathcal{F}_a^{(n)}(\bfk_1,\cdots,\bfk_n)=\sum_{m=1}^{n-1}\sigma_{ab}(n)\,\gamma_{bcd}(\bfk_{1\cdots m},\bfk_{(m+1) \cdots n})\,\mathcal{F}_c^{(m)}(\bfk_1,\cdots,\bfk_m) \mathcal{F}_d^{(n-m)}(\bfk_{m+1},\cdots,\bfk_n)
\label{eq:PT_recursion_fourier}
\end{align}
with the initial condition, $\mathcal{F}_a^{(1)}\equiv(1,1)$. The explicit functional form of $\sigma_{ab}$ and $\gamma_{bcd}$ can be found in e,g., Refs.~\cite{Goroff:1986ep, Bernardeau:2001qr, Crocce:2005xy, Nishimichi:2007xt}. One advantage of the Fourier-space formulation in Eqs.~(\ref{eq:delta_n_Fourier})-(\ref{eq:PT_recursion_fourier}) is that cosmology dependence are entirely separated out, and the structure of the PT kernels is determined irrespective of the initial conditions and background cosmology (but see Refs.~\cite{Taruya2016,Bose_Koyama2016,Fasiello_Vlah2016} for generalized cosmologies). However, if one wants to compare the PT prediction with a particular realization of the $N$-body simulation and/or observed large-scale structure at field level,  the Fourier-space formulation \cite{roth/porciani:2011} becomes impractical to evaluate the higher-order density and velocity fields because of the multi-dimensional convolution integrals. We therefore implement the right hand side of Eq.~(\ref{eq:recursion_formula}) directly in the real space.

\subsection{{\tt GridSPT}: Generating higher-order density fields on grids}
\label{sec:gridSPT}

In this subsection, based on the real-space recursion relations, Eqs.~(\ref{eq:recursion_formula}) and (\ref{eq:recursion_n=1}), we present a grid-based PT calculation, called GridSPT, which enables us to systematically evaluate the higher-order PT solutions at the field level. Making use of the Fast-Fourier Transform (FFT), the basic procedure to construct the density and velocity fields on grids are as follows. Note that 
one can find in the literature some studies closely related to this work, in which the FFT technique has been applied to directly solve Eqs.~(\ref{eq:eq_continuity})-(\ref{eq:eq_Poisson}) \cite{Gouda1995,Gouda1994}. Here, we rather stick to a perturbative calculation, and give a recipe to compute the PT solutions on grids: 

\begin{enumerate}
\renewcommand{\labelenumi}{(\arabic{enumi}).}
\item Generate the initial density field $\delta_0(\bfk)$ on Fourier-space grids drawn from the Gaussian random distribution specified by the linear power spectrum $P_0(k)$. 
\item 
Perform the inverse FFT to obtain the following quantities on real-space grids after multiplying the relevant factors to $\delta_0(\bfk)$ in Fourier space: 
\begin{align}
\left.
\begin{array}{l}
\delta_0(\bfk)
\\
\\
i\bfk\,\delta_0(\bfk)
\\
\\
{\displaystyle \Bigl(-\frac{i\bfk}{k^2}\Bigr)\delta_0(\bfk) }
\\
\\
{\displaystyle \Bigl(\frac{k_ik_j}{k^2}\Bigr) \delta_0(\bfk) }
\end{array}
\right\}
\stackrel{{\rm\scriptscriptstyle inverse\,\,FFT}}{\Longrightarrow}
\left\{
\begin{array}{l}
\delta_1(\bfx)=\theta_1(\bfx)
\\
\\
\nabla\delta_1(\bfx)=\nabla\theta_1(\bfx)
\\
\\
\bfu_1(\bfx)
\\
\\
\partial_i(\bfu_1)_j
\end{array}
\right.
\end{align}

\item Substitute the quantities obtained in the step (2) to the recursion relation (\ref{eq:recursion_formula}) to construct the second-order solutions $\delta_2(\bfx)$ and $\theta_2(\bfx)$ on real-space grids.

\item Perform the forward FFT to have $\delta_2(\bfk)$ and $\theta_2(\bfk)$. 

\item Perform the inverse FFT to obtain the following quantities on real-space grids:
\begin{align}
\left.
\begin{array}{l}
i\bfk\,\delta_2(\bfk)
\\
\\
i\bfk\,\theta_2(\bfk)
\\
\\
{\displaystyle \Bigl(-\frac{i\bfk}{k^2}\Bigr)\theta_2(\bfk) }
\\
\\
{\displaystyle \Bigl(\frac{k_ik_j}{k^2}\Bigr) \theta_2(\bfk) }
\end{array}
\right\}
\stackrel{{\rm\scriptscriptstyle inverse\,\,FFT}}{\Longrightarrow}
\left\{
\begin{array}{l}
\nabla\delta_2(\bfx)
\\
\\
\nabla\theta_2(\bfx)
\\
\\
\bfu_2(\bfx)
\\
\\
\partial_i(\bfu_2)_j
\end{array}
\right.
\end{align}

\item Use the recursion relation and quantities obtained so far (i.e., $\delta_m(\bfx)$, $\theta_m(\bfx)$, $\nabla\delta_m$, $\nabla\theta_m$, $\bfu_m$, and $\partial_i(\bfu_m)_j$ for $m=1$ and $2$) to construct the third-order solutions $\delta_3(\bfx)$ and $\theta_3(\bfx)$ on real-space grids. 
\end{enumerate}

\bigskip
Repeating the last three steps $(4)\sim(6)$, this procedure can be generalized to an arbitrary higher-order order in PT solutions. That is, provided the solutions up to the $n$-th order, the $(n+1)$-th order solutions, $\delta_{n+1}$ and $\theta_{n+1}$, are constructed using FFT. In this paper, we will explicitly demonstrate the grid-based PT calculations up to fifth order. Note that in computing $(n+1)$-th order, we do not necessarily store all the field data set up to $n$-th order. What is needed is the density and velocity-divergence data up to the $n$-th order, from which we can generate the gradient and vector/tensor fields in Fourier space.

Although the above procedure is simple and thus the code implementation is rather straightforward, one needs to take care of the aliasing effect in practice. The aliasing effect arises when we Fourier-transform the nonlinear terms in the right hand side of Eq.~(\ref{eq:recursion_formula}). For simplicity, consider the two fields, $a(x)$ and $b(x)$, in one-dimensional grid space of $-L/2\leq x\leq L/2$ with a grid number $N$. The discrete Fourier transform is described by
\begin{align}
 a(x_j)=\sum_{n=-N/2}^{N/2-1} a(k_n)\,e^{i k_n\,x_j},\quad a(k_n)=\frac{1}{N}\sum_{j=-N/2}^{N/2-1} a(x_j)\,e^{-i\,k_n x_j}
\end{align}
with $k_n=2n\pi/L$ and $x_j=(j/N)L$ for $j=-N/2,\cdots, N/2-1$. Then the Fourier transform of the product, $c(x)\equiv a(x)b(x)$ , leads to
\begin{align}
 c(k_n) = \frac{1}{N}\sum_{j=-N/2}^{N/2-1}c(x_j)\,e^{-i\,k_n x_j}
 = \sum_{l,m=-N/2}^{N/2-1} \delta^{\rm K}_{l+m,n}\,a(k_l)b(k_m) + 
\sum_{l,m=-N/2}^{N/2-1} \delta^{\rm K}_{l+m,n\pm N}\,a(k_l)b(k_m)
\label{eq:product_Fourier}
\end{align}
with $\delta^{\rm K}$ being the Kronecker delta. 
In the right hand side of Eq.~(\ref{eq:product_Fourier}), the first term is the signal that we want to calculate, and second term is the aliasing contribution coming from the discrete sampling. A simple way to eliminate the aliasing effect is to discard the high-frequency modes that produces the aliasing contribution. For example, if we force the fields $a(k_n)$ and $b(k_n)$ to zero at $|n|> N/3$, the non-vanishing modes that appear in Eq.~(\ref{eq:product_Fourier}) are restricted to $|l+m-n|<N$, and hence the aliasing term never appears. This zero-padding method, referred to as the $2/3$ rule, can be also applied to the three-dimensional case (e.g., \cite{Orszag1971a}). The method suits well for our purpose because the validity of the PT calculation is basically limited at large-scale modes (i.e., low-$k$ modes).

In what follows, we will demonstrate the grid-based PT calculations with side length of the boxsize $L_{\rm box}=1,000\,h^{-1}$Mpc and number of grids $N_{\rm grid}=512^3$. We then eliminating the aliasing effect by applying the isotropic low-pass  filter (called sharp-$k$ filter) in the Fourier space to all variables. Note that strict $2/3$ rule would suggest to set the modes with $|k_{x,y,z}|>k_{\rm crit}\equiv(2\pi/L_{\rm box})(N_{\rm grid}^{1/3}/3)\simeq 1.07\,h$\,Mpc$^{-1}$ to zero. After carefully investigating the impact of cutoff scales on the statistical results in Appendix \ref{Appendix:cutoff_depdendence}, we find that employing the two different cutoff wavenumbers performs better. As the default setup, we apply the sharp-$k$ filter of $k_{\rm cut,1}=1\,h$\,Mpc$^{-1}$ to the linear density fields, and then apply the same sharp-$k$ filter with $k_{\rm cut,2}=(4/3)\,h$\,Mpc$^{-1}$ to the higher-order density fields. Though this exceeds the critical wavenumber $k_{\rm crit}$, the resultant PT calculations give the best performance. A possible reason for this is discussed in Appendix \ref{Appendix:cutoff_depdendence}. 

Implementing this filtering technique, the code of grid-based PT calculations, which we hereafter call {\tt GridSPT}, is written in {\tt c++} using FFTW library\footnote{version 2.1.5; {\tt http://www.fftw.org}}. With the thread-parallerized FFT calculation, the code can quickly generates higher-order density fields\footnote{To be precise, with the CPU of Xeon E5-2695 2.1GHz (36 cores) and using the Intel compiler and 36 threads for FFT, it takes roughly two minutes to generate density fields up to fifth order.}.   

\section{Properties of SPT density fields}
\label{sec:demonstration}

Here, we present the results of grid-based PT calculations up to fifth order, and in comparison with $N$-body simulations, we investigate in detail the properties of generated density fields. The cosmological parameters used to generate linear density field as well as to run the $N$-body simulation are determined by WMAP5 results \cite{Komatsu:2008hk} assuming the flat-$\Lambda$CDM model: $\Omega_{\rm m}=0.279$ for matter density, $\Omega_{\Lambda}=0.721$ for dark energy with equation-of-state parameter $w=-1$, $\Omega_{\rm b}/\Omega_{\rm m}=0.165$ for baryon fraction, $h=0.701$ for Hubble parameter, $n_s=0.96$ for scalar spectral index, and finally, $\sigma_8=0.8159$ for the normalization of the fluctuation amplitude at $8\,h^{-1}$Mpc. The cosmological $N$-body simulation is carried out by publicly available code, GADGET-2 \cite{Springel:2005mi}, with $N_{\rm particle}=1,024^3$ particles in comoving periodic cubes of $L_{\rm box}=1,000\,h^{-1}$Mpc, starting with  the initial density field calculated from the {\tt 2LPT} code \cite{Crocce:2006ve} at redshift $z_{\rm init}=30$. Note that we use the same initial density field ($512^3$ large-scale modes) for the {\rm GridSPT} calculation.

\begin{figure}[h]
\vspace*{-0.8cm}
\begin{center}
\includegraphics[width=14.5cm,angle=0]{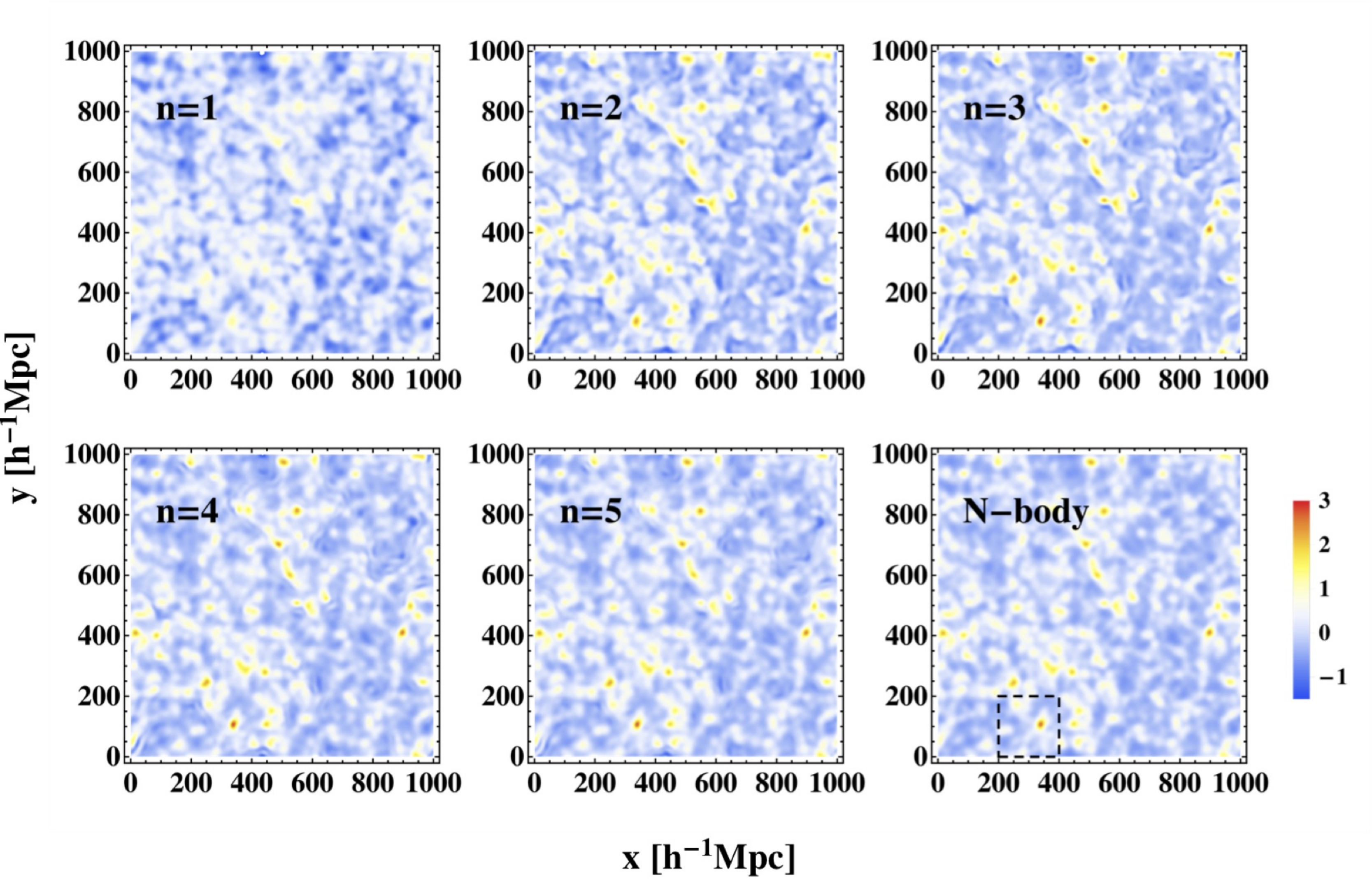}
\end{center}
\vspace*{-0.6cm}
\caption{2D density field at $z=0$ smoothed with Gaussian filter of $R=10\,h^{-1}$Mpc. The results generated with {\tt GridSPT} code are shown (from top left to bottom middle), averaging over the  $10\,h^{-1}$Mpc depth in each grid. Here, the color scale represents the amplitude of the density field, $\delta_{\rm SPT}=\sum_{i=1}^n\,D_+^i\,\delta_i$ with the number $n$ indicated in each panel. For comparison, bottom right panel shows the density field from $N$-body simulation, evolved with the same initial condition as used in {\tt GridSPT} calculations.   
\label{fig:Slice_FullSize}
}
\vspace*{-0.7cm}
\begin{center}
\includegraphics[width=14.5cm,angle=0]{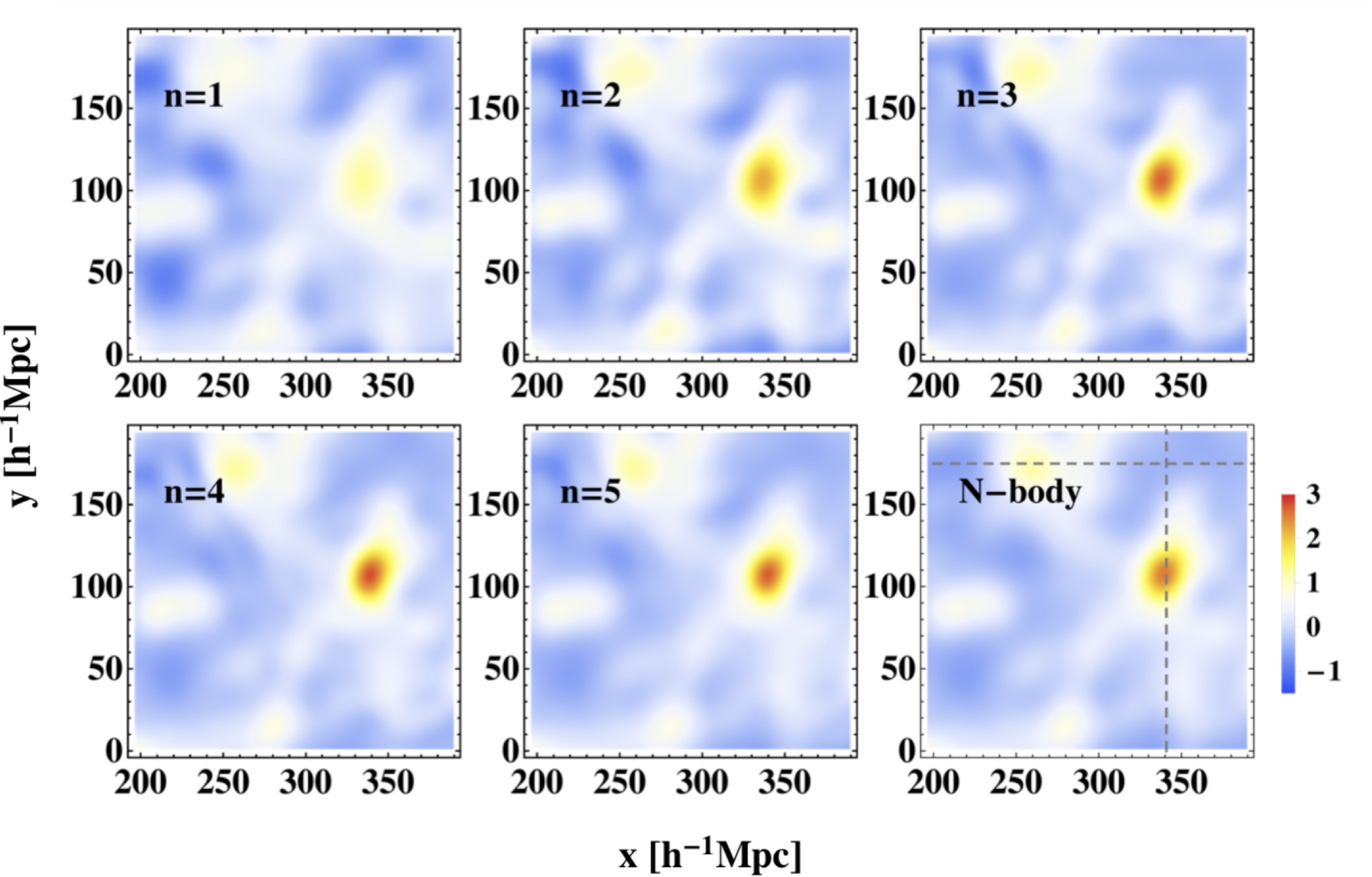} 
\end{center}
\vspace*{-0.6cm}
\caption{Same as in Fig.~\ref{fig:Slice_FullSize}, but zoom-in plot of the 2D density field over $200\times200\,h^{-1}$Mpc size is particularly shown for the region enclosed by the dashed line in bottom right panel of Fig.~\ref{fig:Slice_FullSize}. 
\label{fig:Slice_ZoomIn}
}
\end{figure}

\subsection{The morphology of non-linear density fields}
\label{subsec:structure}

Let us first present the morphology of density fields generated by {\tt GridSPT}, and investigate their properties, comparing with $N$-body results. Figs.~\ref{fig:Slice_FullSize} and \ref{fig:Slice_ZoomIn} show the 2D slices of the density fields at $z=0$ from {\tt GridSPT} (with the order $n$ label) and from $N$-body results (bottom right). In Fig.~\ref{fig:Slice_FullSize}, the density fields over the entire box are shown with the amplitude plotted in linear scale. On the other hand, Fig.~\ref{fig:Slice_ZoomIn} shows the local density fields particularly taken from Fig.~\ref{fig:Slice_FullSize} over the $200\times200\,h^{-1}$Mpc-sized region enclosed by dashed line in bottom right panel. In both figures, we apply the Gaussian filter of the radius $R=10\,h^{-1}$Mpc to the results, and on each grid, the density fields are averaged over the  $10\,h^{-1}$Mpc depth. While the top left panel in Figs.~\ref{fig:Slice_FullSize} and \ref{fig:Slice_ZoomIn} are the linear density field (labeled as $n=1$), the four successive panels (i.e., top middle, top right, bottom left, and bottom middle) are the \texttt{GridSPT} results summing up higher-order density fields, i.e., $\delta_{\rm SPT}=\sum_{i=1}^n\,D_+^i\,\delta_i$ with the number $n$ indicated in each panel. For comparison, the last panel (bottom right) is the $N$-body result obtained from the same initial density field. Adding higher-order PT corrections, the resultant density fields tend to show a clearer contrast, and resemble the $N$-body density field. At the fifth order, the PT density field seems to almost match the $N$-body result from a visual inspection. It is indeed hard from Fig.~\ref{fig:Slice_FullSize} to discriminate between the N-body and {\tt GridSPT}, but a closer look at high-density region seen in Fig.~\ref{fig:Slice_ZoomIn} reveals that there is a slight mismatch between the PT prediction (at the fifth order) and the simulation result.

\begin{figure*}[tb]
\vspace*{-1.5cm}
\begin{center}
\includegraphics[width=16cm,angle=0]{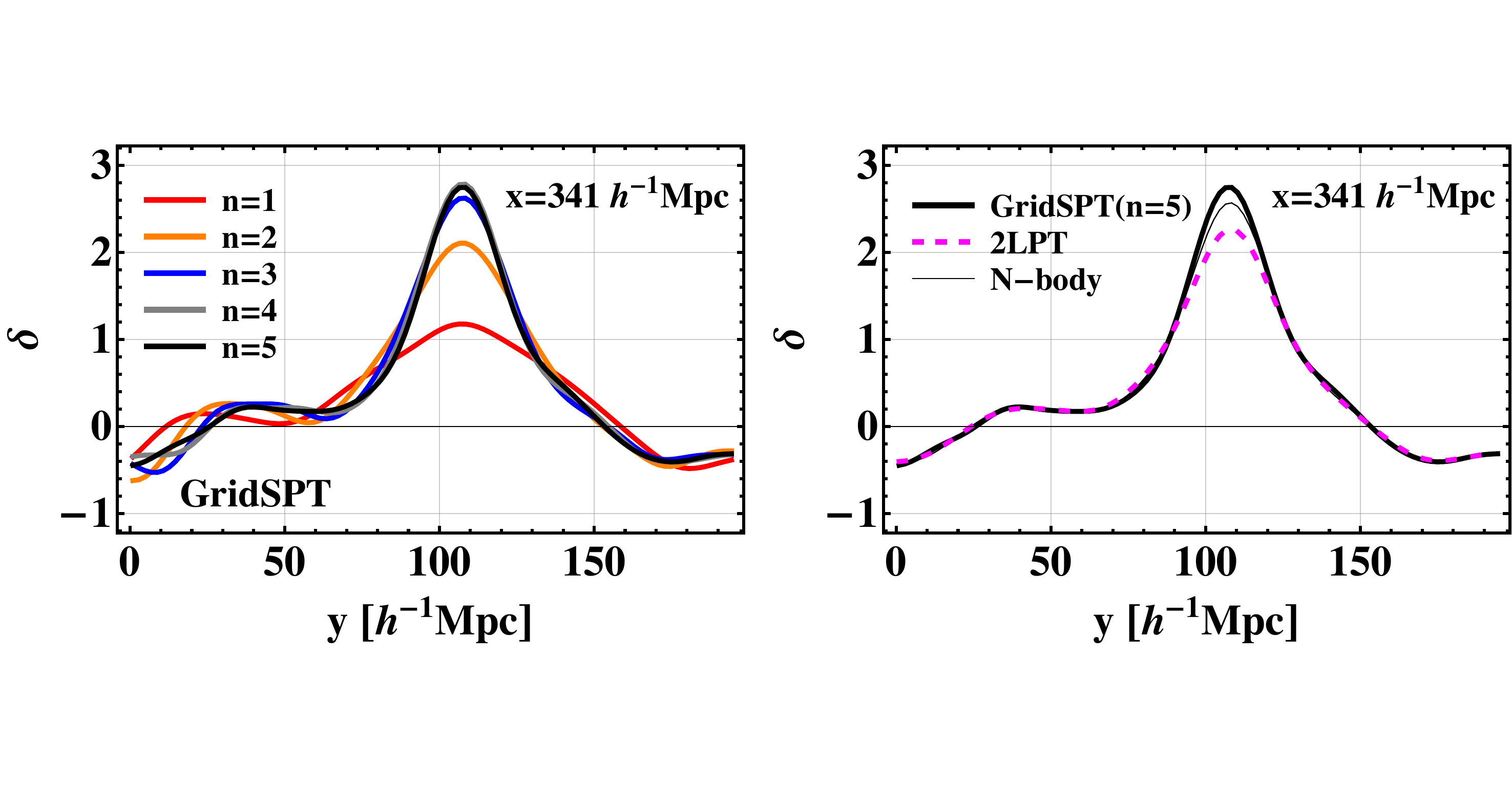}
\end{center}
\vspace*{-1.9cm}
\caption{1D density field at $z=0$ smoothed with Gaussian filter of $R=10\,h^{-1}$Mpc, taken from Fig.~\ref{fig:Slice_ZoomIn}. The density field shown here lies at $x=341\,h^{-1}$\,Mpc, indicated as vertical dashed line in bottom right panel of Fig.~\ref{fig:Slice_ZoomIn}.
\label{fig:Slice_1D_high}
}
\vspace*{-1.3cm}
\begin{center}
\includegraphics[width=16cm,angle=0]{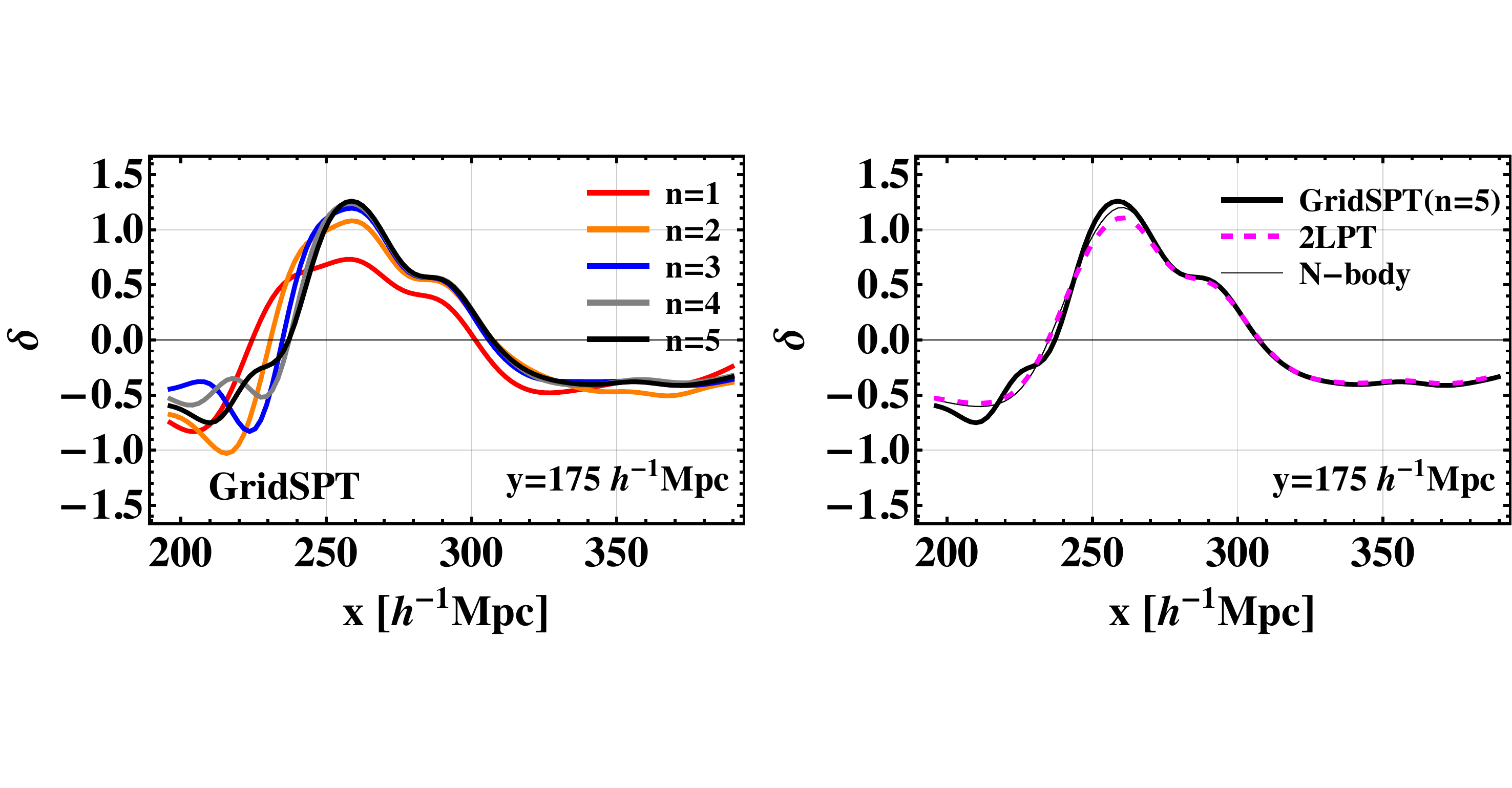}
\end{center}
\vspace*{-1.9cm}
\caption{Same as in Fig.~\ref{fig:Slice_1D_high}, but the low-density region at $y=175\,h^{-1}$\,Mpc is particularly shown (indicated as horizontal dashed line in bottom right panel of Fig.~\ref{fig:Slice_ZoomIn}).
\label{fig:Slice_1D_low}
}
\end{figure*}

To scrutinize the difference between the two, we select representative regions from the local patch in Fig.~\ref{fig:Slice_ZoomIn} (indicated in horizontal and vertical dashed lines), and plot the 1D density fields in Figs.~\ref{fig:Slice_1D_high}  and \ref{fig:Slice_1D_low}. The left panels plot the \texttt{GridSPT} density fields, $\delta_{\rm SPT}=\sum_{i=1}^n\,D_+^i\delta_i$ with $n=1-5$ (indicated in the panel), while the right panels compare the result of $n=5$ with those obtained from the $N$-body simulation and second-order Lagrangian PT with {\tt 2LPT} code. As increasing the order of perturbative expansion, the density peak becomes steepened, and the amplitude gets increased at high density regions, and the low-density regions are flattened conversely. A notable point may be that the convergence of the PT predictions tends to be slow at low density regions. This implies that the PT prediction is generally poor to describe the bulk matter flows, consistent with what has been found in Ref.~\cite{roth/porciani:2011,Tassev2014}. On the other hand, the convergence at high-density regions looks faster, and at the fifth order, the PT prediction reproduces the $N$-body density fields remarkably well. Although there is a general trend that the predicted peak amplitude slightly overestimates the $N$-body results, the overall structure of density peaks is better described by the fifth-order SPT than second-order Lagrangian PT. Nevertheless, this does not imply that the statistical correlation of SPT with $N$-body density field is better than that of Lagrangian PT. This point will be discussed in detail in Sec.~\ref{sec:pkcross} and \ref{sec:jointPDF}.

\begin{figure*}[tb]
\vspace*{-1.0cm}
\begin{center}
\includegraphics[width=16cm,angle=0]{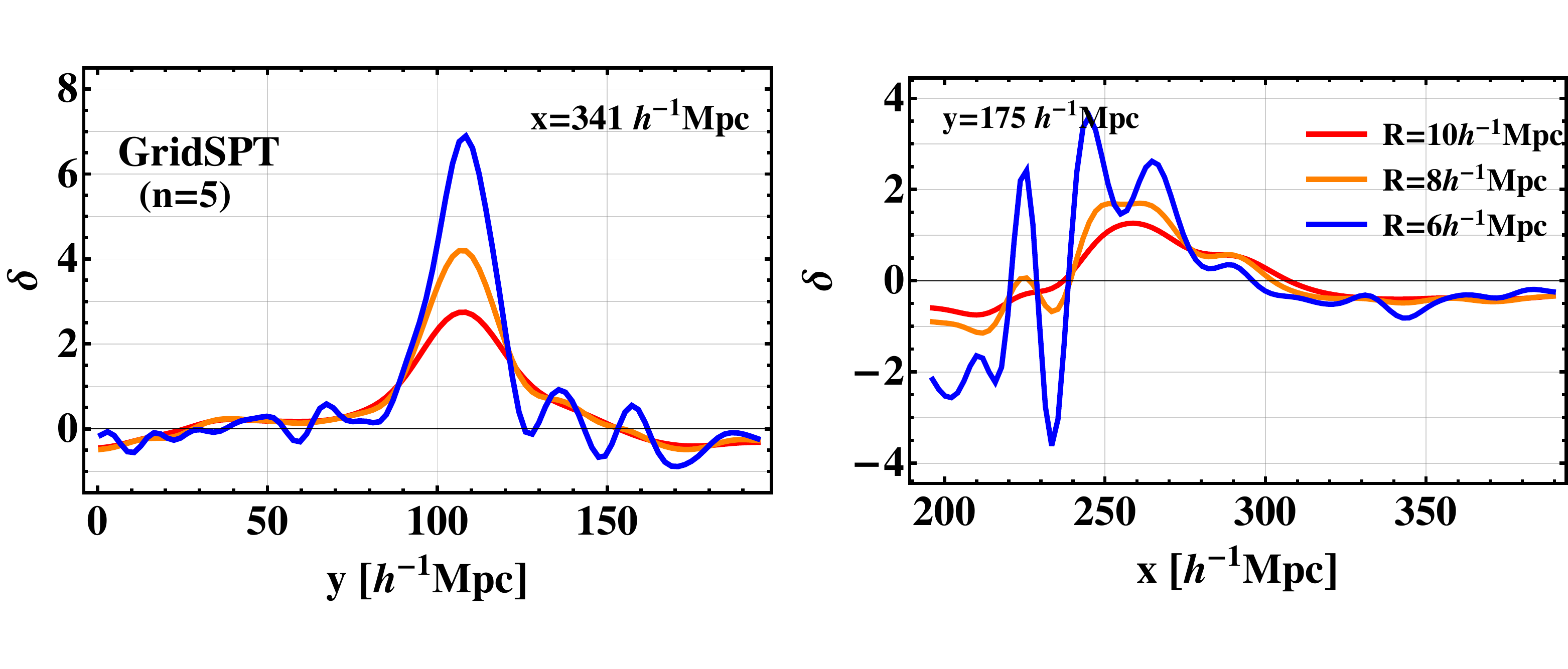}
\end{center}
\vspace*{-1.0cm}
\caption{Dependence of smoothing scale on the projected 1D density fields shown in Figs.~\ref{fig:Slice_1D_high} (left) and \ref{fig:Slice_1D_low} (right).
\label{fig:Slice_1D_Rdept}
}
\end{figure*}

Finally, we note here that the resultant density fields obtained from the \texttt{GridSPT} calculation is rather sensitive to the choice of the smoothing scale. Fig.~\ref{fig:Slice_1D_Rdept} plots the 1D density fields at the same regions as shown in Figs.~\ref{fig:Slice_1D_high} and \ref{fig:Slice_1D_low}, but the fifth-order PT results at different smoothing scales are depicted as different colors. A slight decrease of the smoothing scale results in a spurious wiggle structure at low density region, and this can also affect small density peaks, leading to an un-physical behavior of $\delta<-1$. We have checked that this behavior appears persistently in the {\tt GridSPT} density fields smoothed with the same scales, regardless of the box size and the aliasing correction. The fact that the density field is less than $-1$ is thus not simply due to the numerical artifact, if any, but rather a drawback of Eulerian perturbation theory; for the region with $\delta<-1$, continuity equation yields the violation of mass conservation, namely, density contrast increases when the velocity field is divergent. The explicit violation of mass conservation exhibits a rather strong mode coupling between long and short modes. We will discuss this UV-sensitive behavior from the statistical point-of-view in the next section.

\subsection{Statistical properties}
\label {subsec:statistics}

In this subsection, based on the density fields generated with {\tt GridSPT} in Sec.~\ref{subsec:structure}, we measure the statistical quantities, and study their properties in comparison with the result of the $N$-body simulation. In particular, we measure the cross-correlation of the SPT density field with $N$-body results, and discuss on how well the PT predictions reproduce the N-body results.

\subsubsection{Power spectrum and bispectrum}
\label{subsec:pk_bk}

\begin{figure*}[tb]
\begin{center}
\includegraphics[width=8.7cm,angle=0]{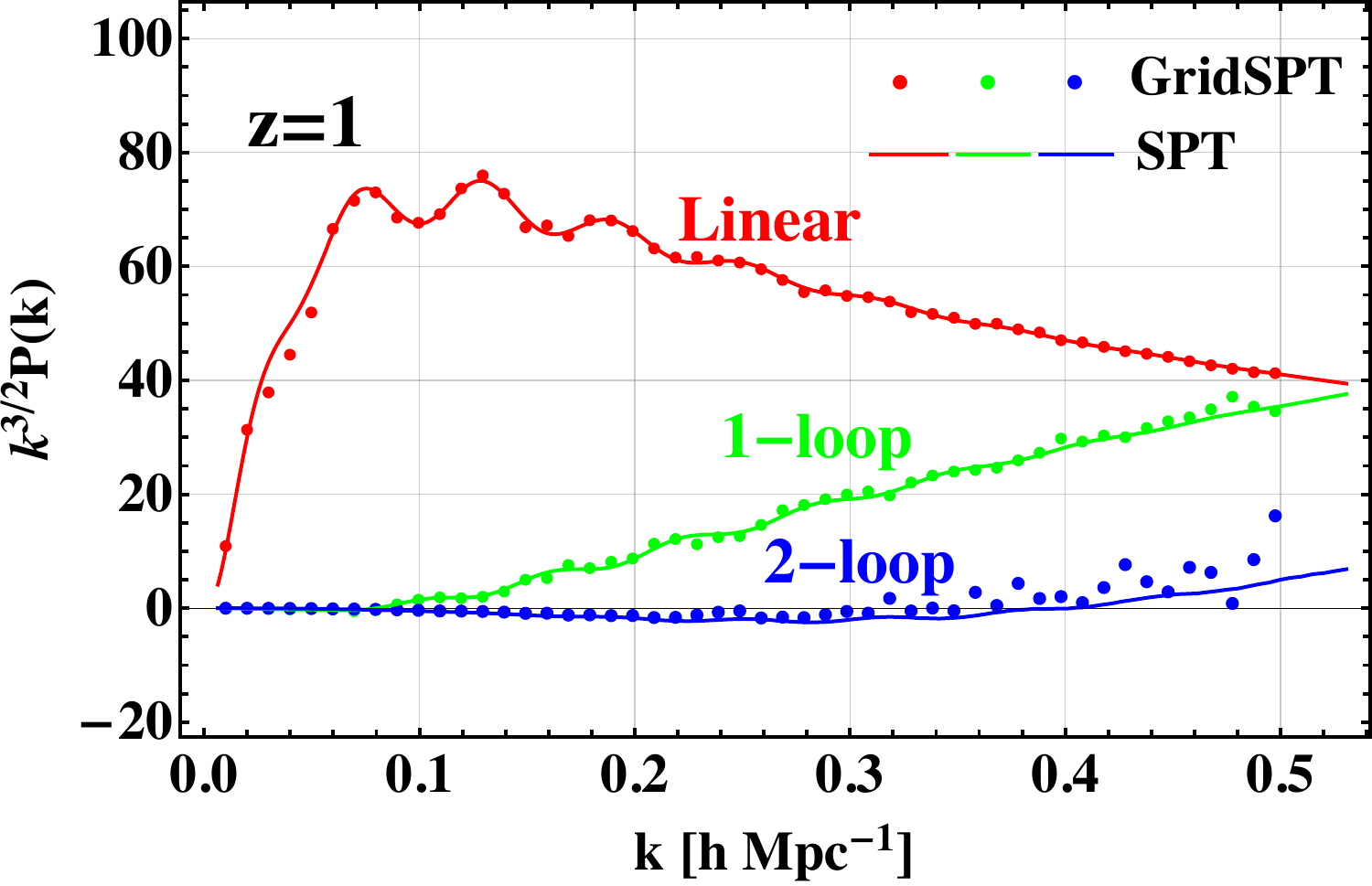}
\hspace*{0.2cm}
\includegraphics[width=8.7cm,angle=0]{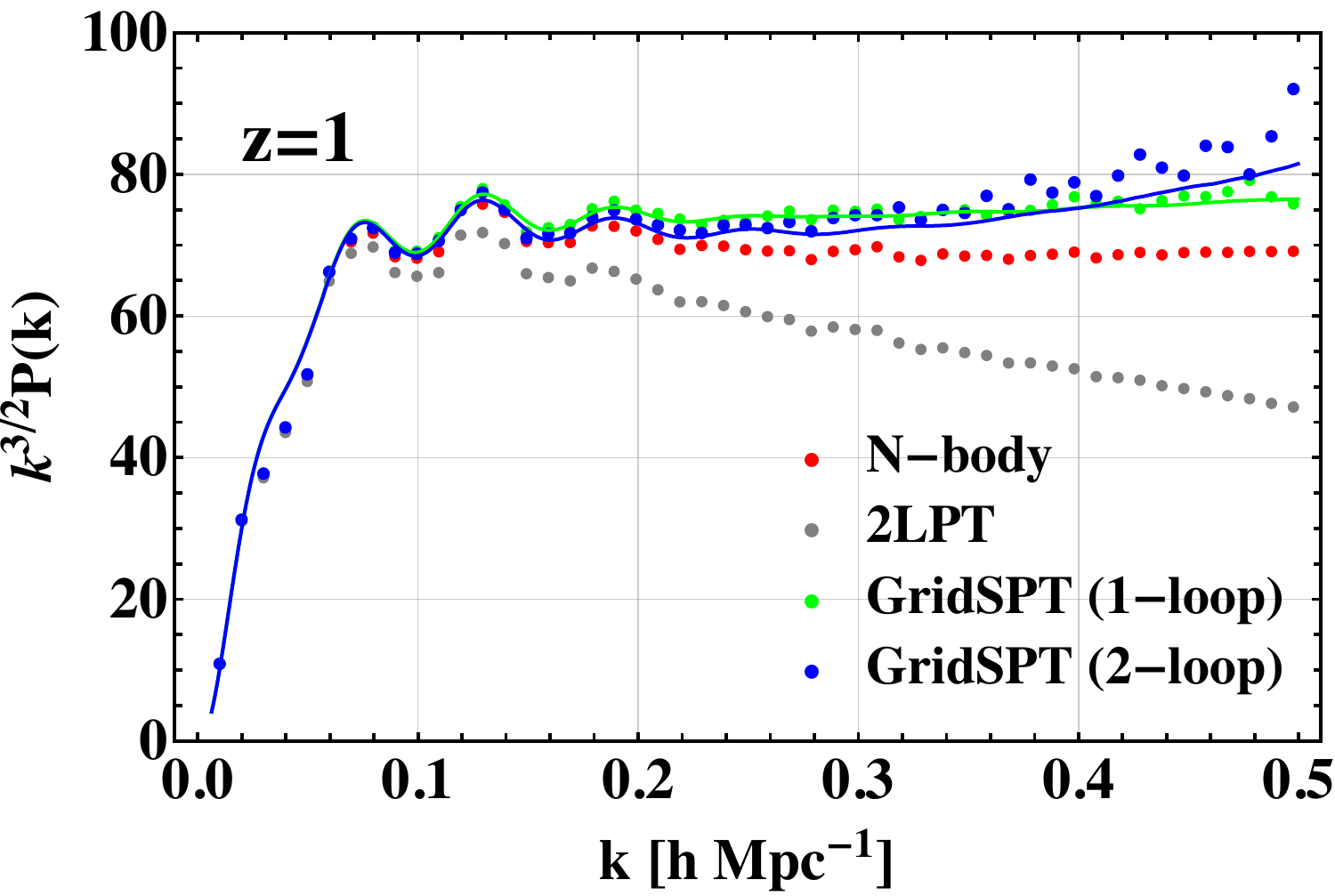}
\end{center}
\vspace*{-0.4cm}
\caption{{\it Left}: Linear power spectrum (red), and one-loop (green) and two-loop (blue) corrections to the power spectrum at $z=1$. The results measured from the {\tt GridSPT} density fields (filled circles) are compared with those obtained from analytic PT calculations (solid). {\it Right}: Comparison of the total power spectrum at $z=1$ between {\tt GridSPT} calculations at one-loop (green) and two-loop (blue) order and $N$-body simulations (red). For comparison, the second-order Lagrangian PT prediction generated with {\tt 2LPT} is also shown in gray filled circles. 
\label{fig:pkdd_1oop_2loop}
}
\end{figure*}

Before discussing the cross-correlation with $N$-body results, let us first check if the {\tt GridSPT} calculations properly reproduce the analytic prediction computed with Fourier-space recursion relation in Eq.~(\ref{eq:PT_recursion_fourier}). Fig.~\ref{fig:pkdd_1oop_2loop} shows the measured results of the power spectrum from the \texttt{GridSPT} calculations, which are compared with both analytic PT predictions and $N$-body results. Here, the results at $z=1$ are particularly shown. All the results are multiplied by $k^{3/2}$ ($k^3$) for the power spectrum (bispectrum). 

In left panel of Fig.~\ref{fig:pkdd_1oop_2loop}, the contribution to the power spectrum at each perturbative order, $P_{\rm lin}$ (red), $P_{\rm 1\mbox{-}loop}$ (green), and $P_{\rm 2\mbox{-}loop}$ (blue), are shown. These are measured from the \texttt{GridSPT} density fields analogously to the standard procedure on the snapshots of $N$-body simulations: 
\begin{align}
P_{\rm lin}(k)& =D_+^2\,\frac{1}{N_k}\sum_{|\bfk|=k}|\delta_1(\bfk)|^2,
\label{eq:pk_lin}
\\
P_{\rm 1\mbox{-}loop}(k)& =D_+^4\,\frac{1}{N_k}\sum_{|\bfk|=k} \Bigl\{ 2\mbox{Re} [\delta_1(\bfk)\delta_3^*(\bfk)]+|\delta_2(\bfk)|^2\Bigr\},
\label{eq:pk_1loop}
\\
P_{\rm 2\mbox{-}loop}(k)& =D_+^6\,\frac{1}{N_k}\sum_{|\bfk|=k} \Bigl\{ 2\mbox{Re} [\delta_1(\bfk)\delta_5^*(\bfk)+\delta_2(\bfk)\delta_4^*(\bfk)]+|\delta_3(\bfk)|^2\Bigr\},
\label{eq:pk_2loop}
\end{align}
with $N_k$ being the number of Fourier modes in a $k$-bin. The corresponding analytic predictions, taking account of the finite box and high-$k$ cutoff\footnote{To be precise, in analytic PT calculations, we introduced the cutoff scales in the linear power spectrum, given by $k_{\rm min}=2\pi/L_{\rm box}\simeq6.28\times10^{-3}\,h$\,Mpc$^{-1}$ and $k_{\rm max}=k_{\rm cut,2}=(4/3)\,h$\,Mpc$^{-1}$. }, are depicted as solid lines. We expect that the \texttt{GridSPT} results would converge to the analytic curves after we take an ensemble average over many different random realizations of the \texttt{GridSPT} density field.

Overall, the measured results reasonably agree with the analytic predictions. However, a closer look at small scales reveals a small discrepancy between \texttt{GridSPT} and analytic calculations in both $P_{\rm 1\mbox{-}loop}$ and $P_{\rm 2\mbox{-}loop}$. In particular, the discrepancy is manifest in the two-loop correction, and \texttt{GridSPT} calculations are prone to overestimate the analytic prediction. This could happen probably due to the accumulation of small numerical flaw at each order in the \texttt{GridSPT} calculations. Indeed, in SPT, higher-order correction of the power spectrum is known to have a heavy cancellation among multiple diagrams at the same order with positive and negative contributions, and the cancellation becomes more significant as we go to higher-order. Thus, even a small error on each diagram at lower order may result in a big systematics at higher-order corrections through an imperfect cancellation.

In right panel of Fig.~\ref{fig:pkdd_1oop_2loop}, summing up all the PT corrections, the \texttt{GridSPT} results are shown in green and blue filled circles, which are compared with analytic PT predictions depicted as solid lines. As anticipated, discrepancy is manifest at two-loop order, while the one-loop results show a tiny amount of error at high-$k$, which apparently looks insignificant. Although this point has to be kept in mind in our subsequent analyses, the discrepancy is large only at the scales where the deviation from the $N$-body result, depicted as filled red circles, is significant. Further, the discrepancy remains mild, and is smaller than a large underestimation found in the prediction based on {\tt 2LPT} (filled gray circles). 

\begin{figure*}[tb]
\begin{center}
\includegraphics[width=8.7cm,angle=0]{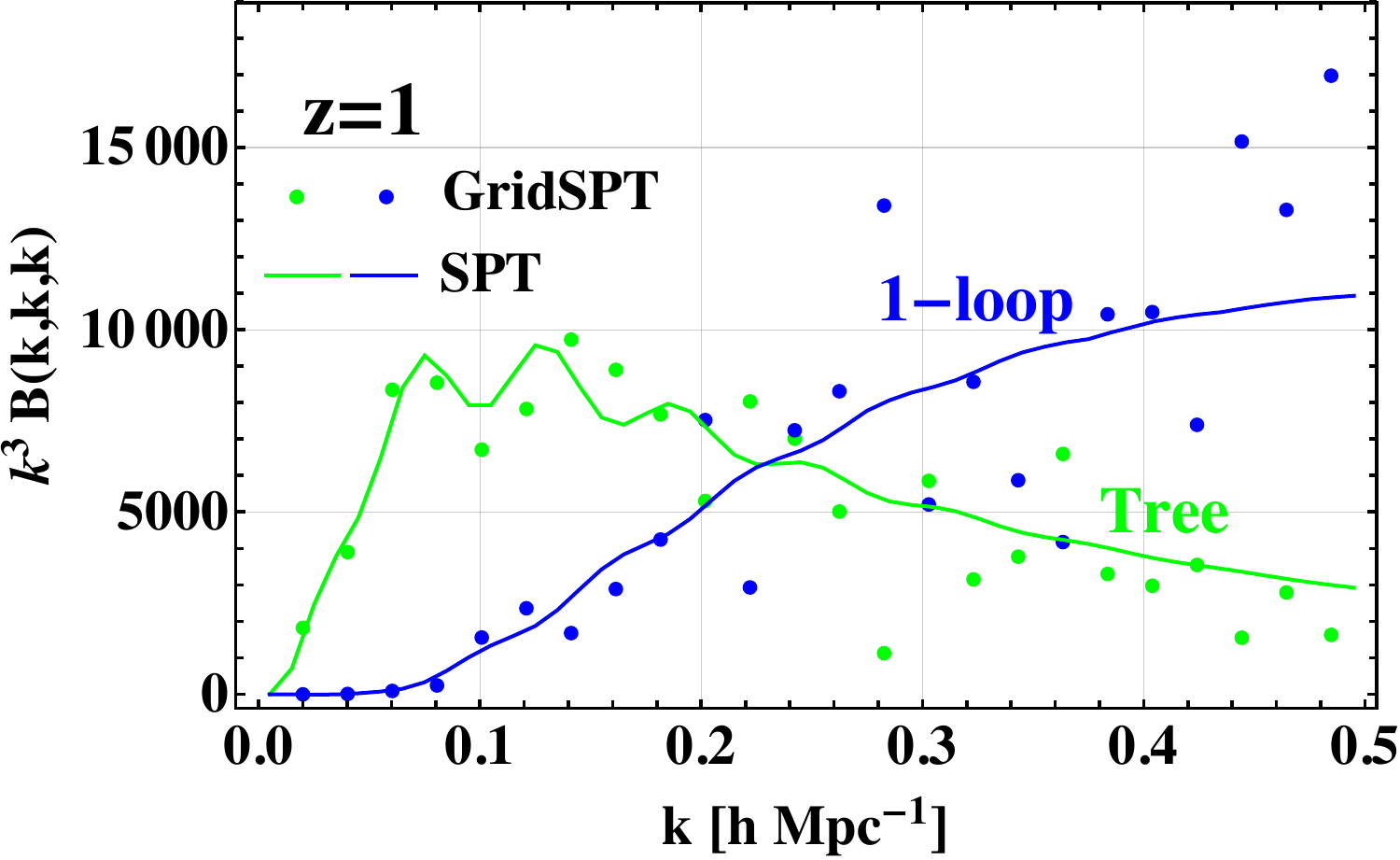}
\hspace*{0.2cm}
\includegraphics[width=8.7cm,angle=0]{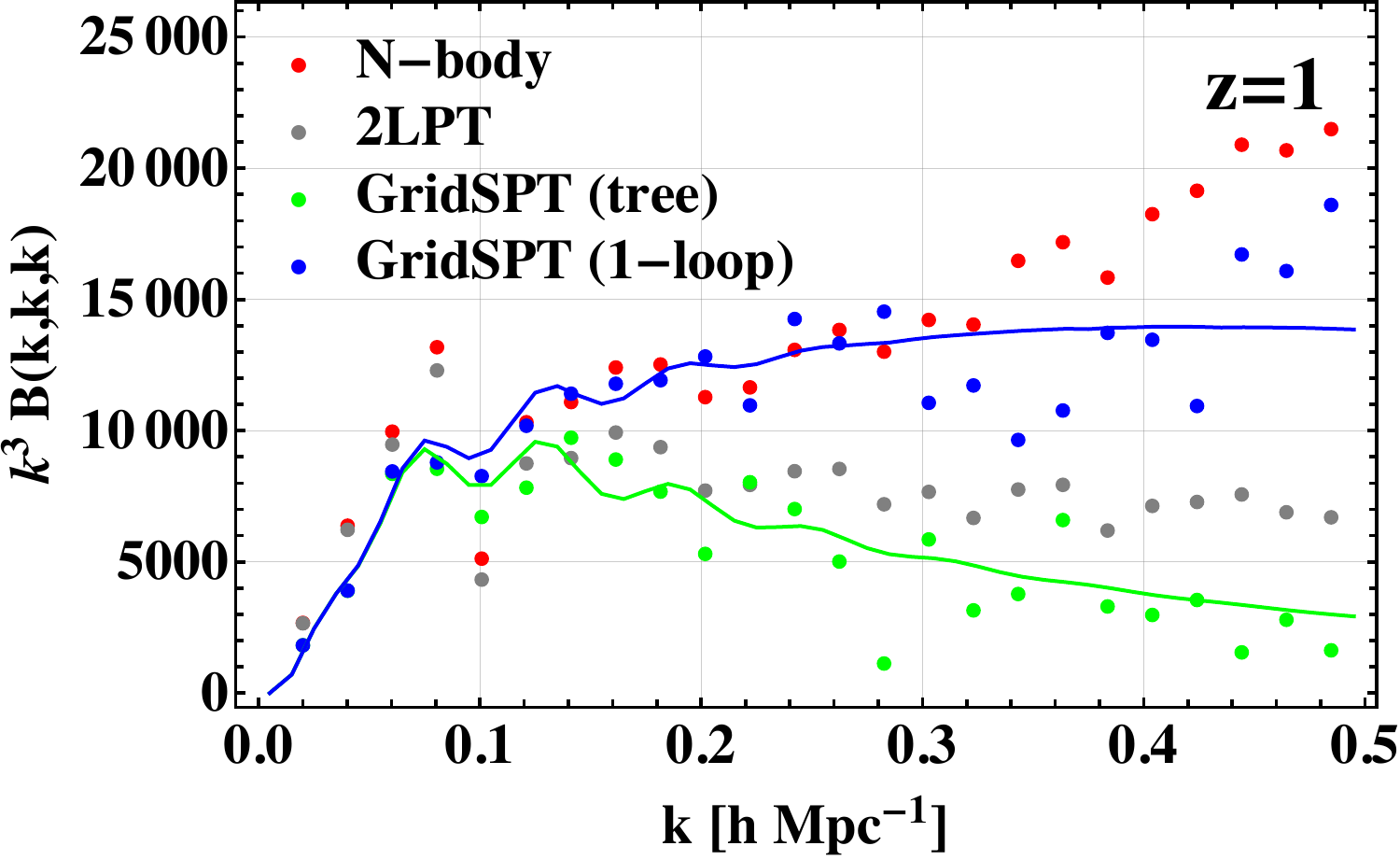}
\end{center}
\vspace*{-0.4cm}
\caption{{\it Left}: Comparison of the tree-level (green) and one-loop (blue) corrections to the bispectrum measured from {\tt GridSPT} density fields (filled circles) with those obtained from analytic PT calculations (solid). The results at equilateral configuration $(k_1=k_2=k_3\equiv k)$ is plotted as function of $k$. {\it Right}: Comparison of the total bispectrum for the equilateral configuration between {\tt GridSPT} calculations at tree-level (green) and one-loop (blue) order and $N$-body simulations (red). The second-order Lagrangian PT prediction generated with {\tt 2LPT} is also shown in gray filled circles. 
\label{fig:bkdd_compared_with_nbody}
}
\vspace*{0.2cm}
\begin{center}
\includegraphics[width=8.7cm,angle=0]{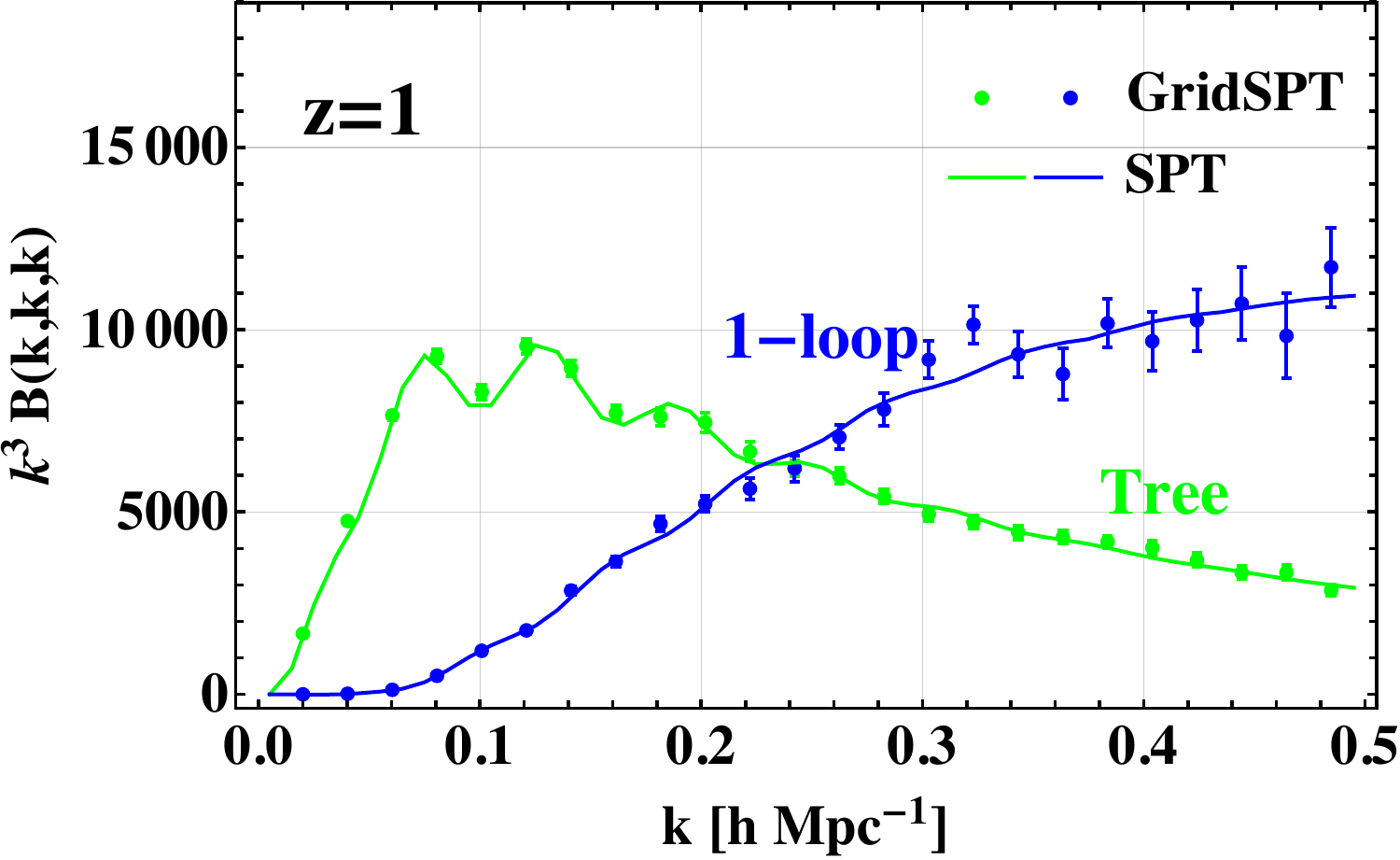}
\hspace*{0.2cm}
\includegraphics[width=8.7cm,angle=0]{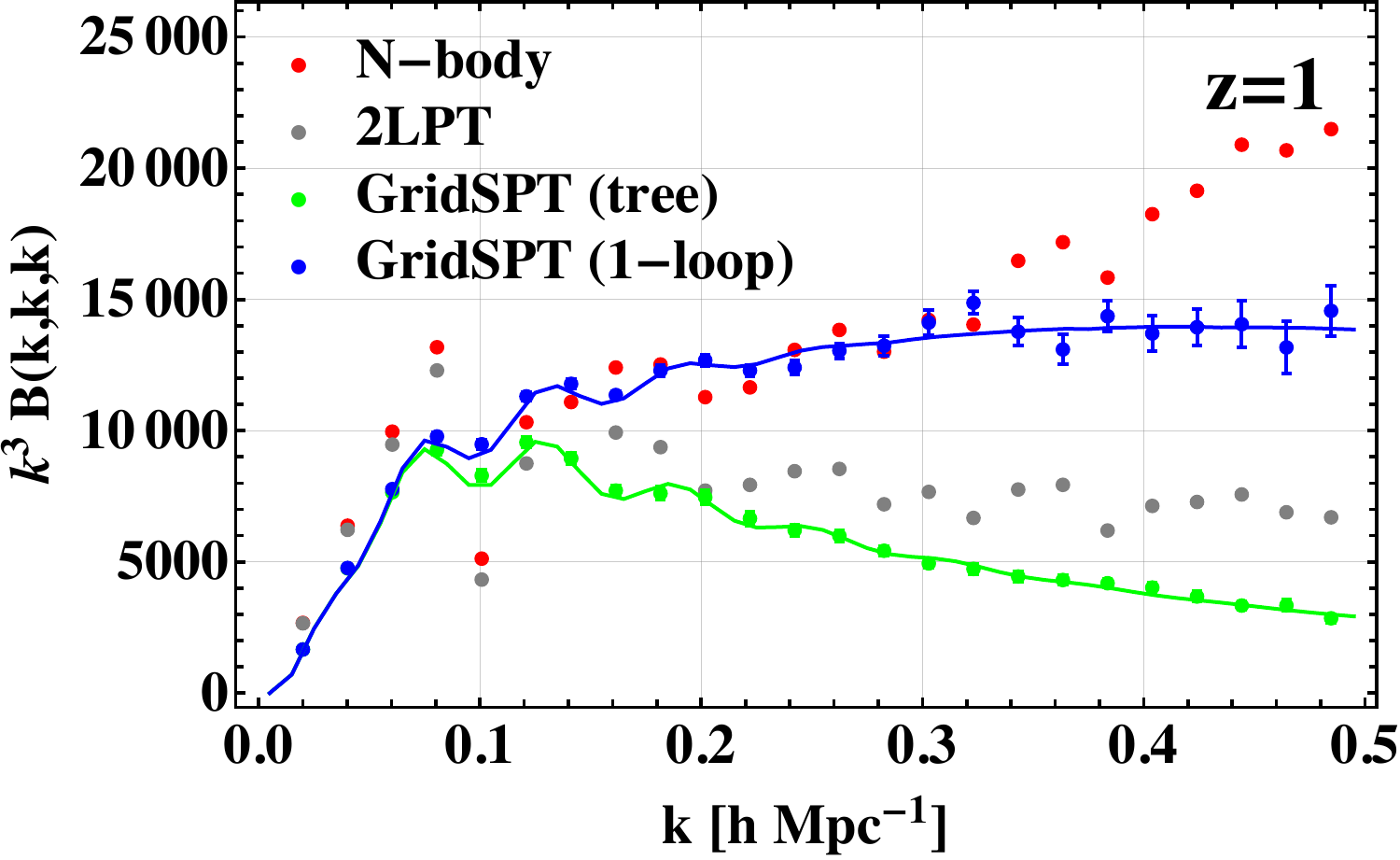}
\end{center}
\vspace*{-0.4cm}
\caption{Same as in Fig.~\ref{fig:bkdd_compared_with_nbody}, but the {\tt GridSPT} results are averaged over $50$ realizations. The errorbars indicate the standard error of the mean. 
\label{fig:bkdd_50realizations}
}
\end{figure*}

Indeed, such a discrepancy is not clearly seen in the case of the bispectrum. Fig.~\ref{fig:bkdd_compared_with_nbody} presents the measured results of the PT contribution at each order, $B_{\rm tree}$ and $B_{\rm 1\mbox{-}loop}$, (left) and their total amplitudes (right), which are compared with $N$-body and {\tt 2LPT} results. Here, the measurement of the bispectrum are done in the equilateral configuration, $k_1=k_2=k_3\equiv k$, and the results are plotted as function of $k$. The PT corrections $B_{\rm tree}$ and $B_{\rm 1\mbox{-}loop}$ are defined as:
\begin{align}
B_{\rm tree}(k_1,k_2,k_3)& =D_+^4\,\frac{1}{N_{\rm 123}}\sum_{\bfk_1,\bfk_2,\bfk_3}\delta^{\rm K}_{\bfk_1+\bfk_2+\bfk_3,\bfzero}\,\,\Bigl[\delta_1(\bfk_1)\delta_1(\bfk_2)\delta_2(\bfk_3)+(\mbox{2 perm.})\Bigr],
\label{eq:bk_tree}
\\
B_{\rm 1\mbox{-}loop}(k_1,k_2,k_3)& =D_+^6\,\frac{1}{N_{\rm 123}}\sum_{\bfk_1,\bfk_2,\bfk_3} \delta^{\rm K}_{\bfk_1+\bfk_2+\bfk_3,\bfzero}\,\,
\Bigl[
\bigl\{\delta_1(\bfk_1)\delta_2(\bfk_2)\delta_3(\bfk_3)+(\mbox{5 perm.}) \bigr\}
\nonumber
\\
&\quad\qquad + \delta_2(\bfk_1)\delta_2(\bfk_2)\delta_2(\bfk_3)+
\bigl\{\delta_1(\bfk_1)\delta_1(\bfk_2)\delta_4(\bfk_3)+(\mbox{2 perm.}) \bigr\}\,
\Bigr].
\label{eq:bk_1loop}
\end{align}
Note that in actual calculation of these expressions, we use a fast estimator based on FFT (e.g., \cite{Baldauf_efatl2015,Sefusatti_etal2016}). Only with the single realization, the resultant bispectrum is rather noisy, however, increasing the number of realizations up to $50$, shown in Fig.~\ref{fig:bkdd_50realizations}, the tree- and one-loop corrections are found to reproduce the analytic PT results (solid lines) quite well. Also, in the right panel, the overall behavior of the one-loop prediction better agrees with $N$-body simulation than the {\tt 2LPT} prediction (gray filled circles).

\begin{figure*}[tb]
\begin{center}
\includegraphics[width=8.7cm,angle=0]{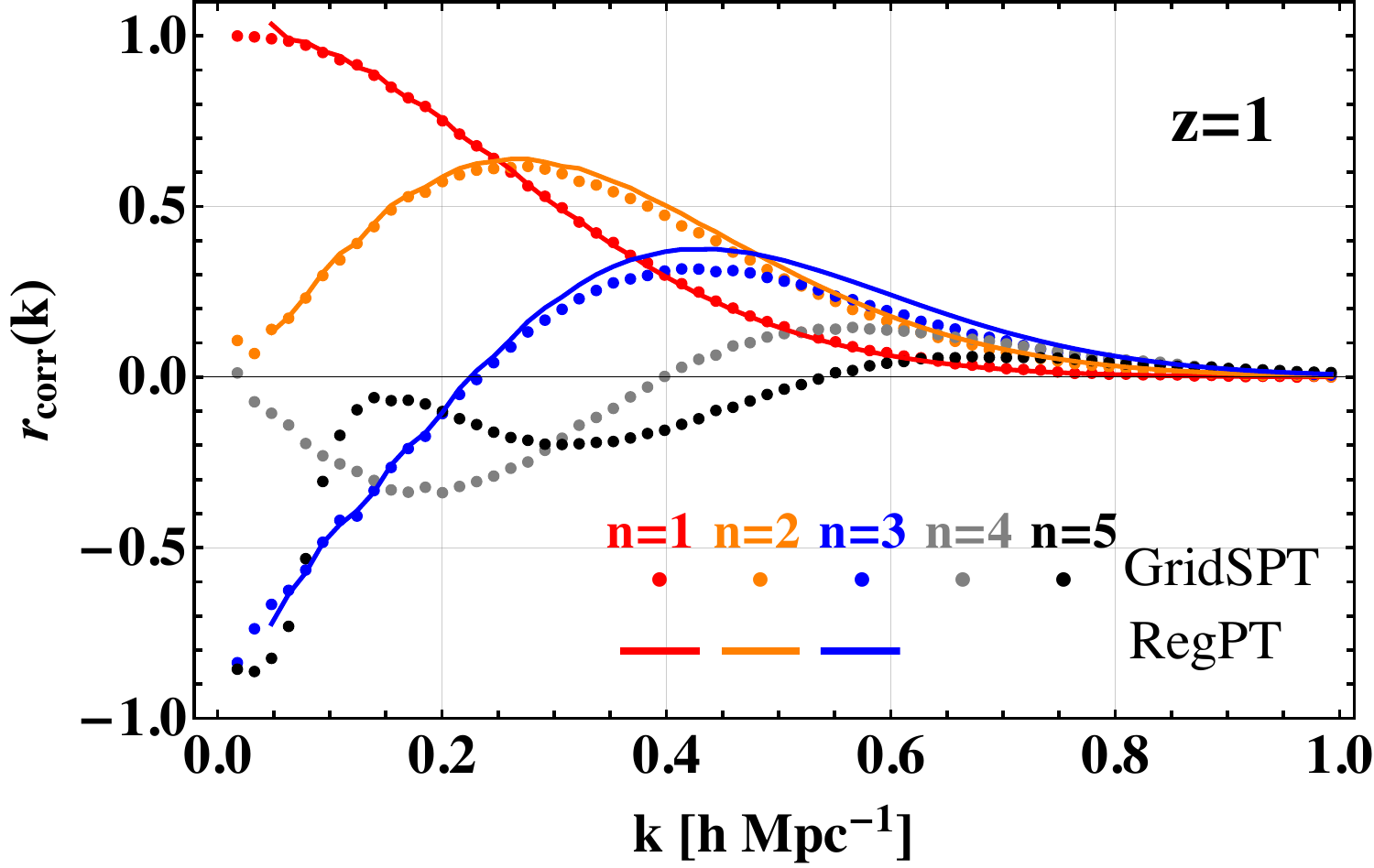}
\hspace*{0.2cm}
\includegraphics[width=8.7cm,angle=0]{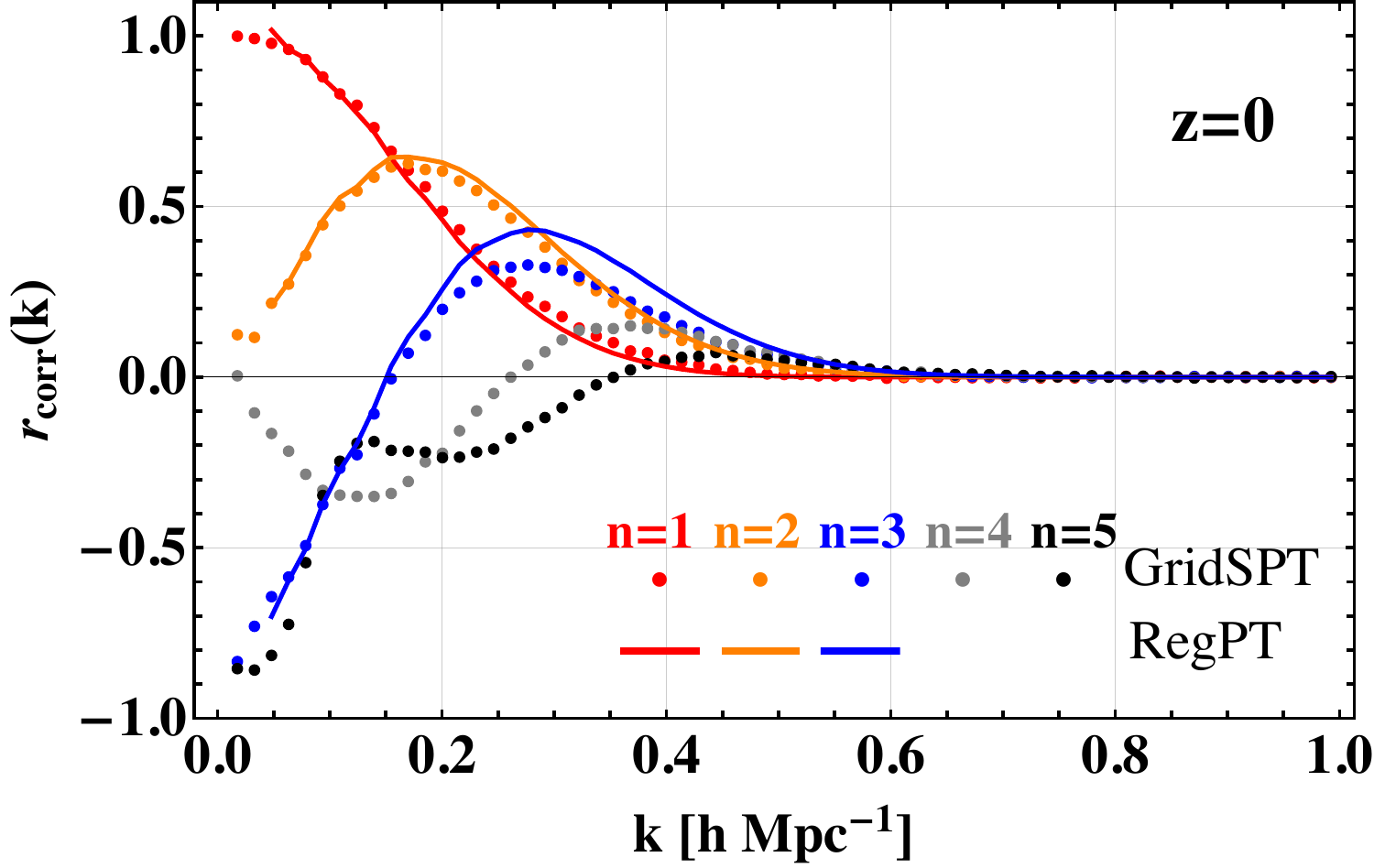}
\end{center}
\vspace*{-0.3cm}
\caption{Cross correlation coefficient for {\tt GridSPT} and $N$-body density fields, $r_{\rm cross}^{(n)}(k)$, defined at Eq.~(\ref{eq:r_corr_nthPT}). The measured results at $z=1$ (left) and $0$ (right) are shown. Solid lines are the predictions based on the RegPT treatment, assuming that the evolved density field in $N$-body simulation is described with the multi-point propagator expansion at two-loop order  (see Appendix \ref{Appendix:pkcross_prediction}). 
\label{fig:Cross_corr_pk_each}
}
\begin{center}
\includegraphics[width=8.7cm,angle=0]{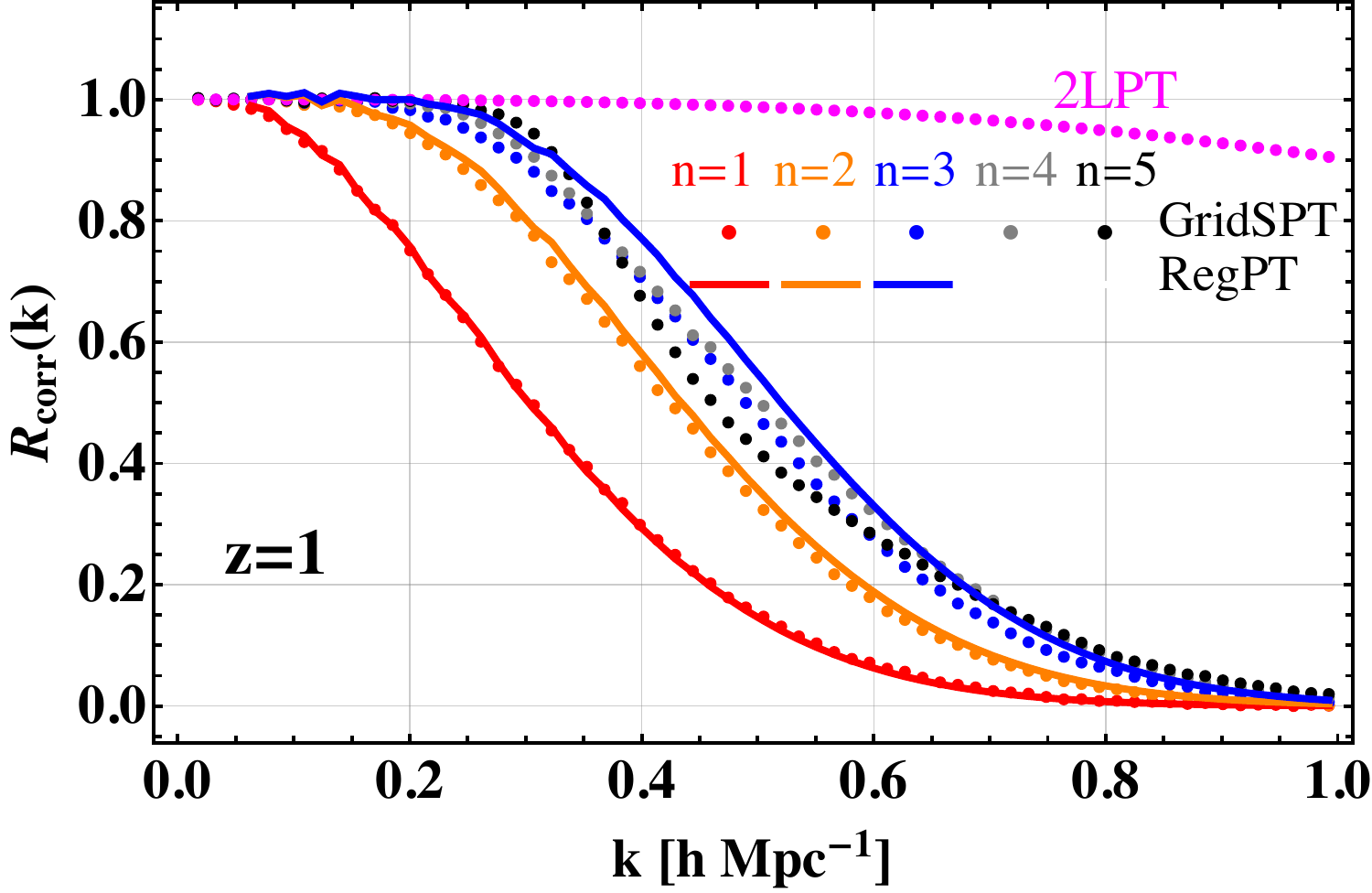}
\hspace*{0.2cm}
\includegraphics[width=8.7cm,angle=0]{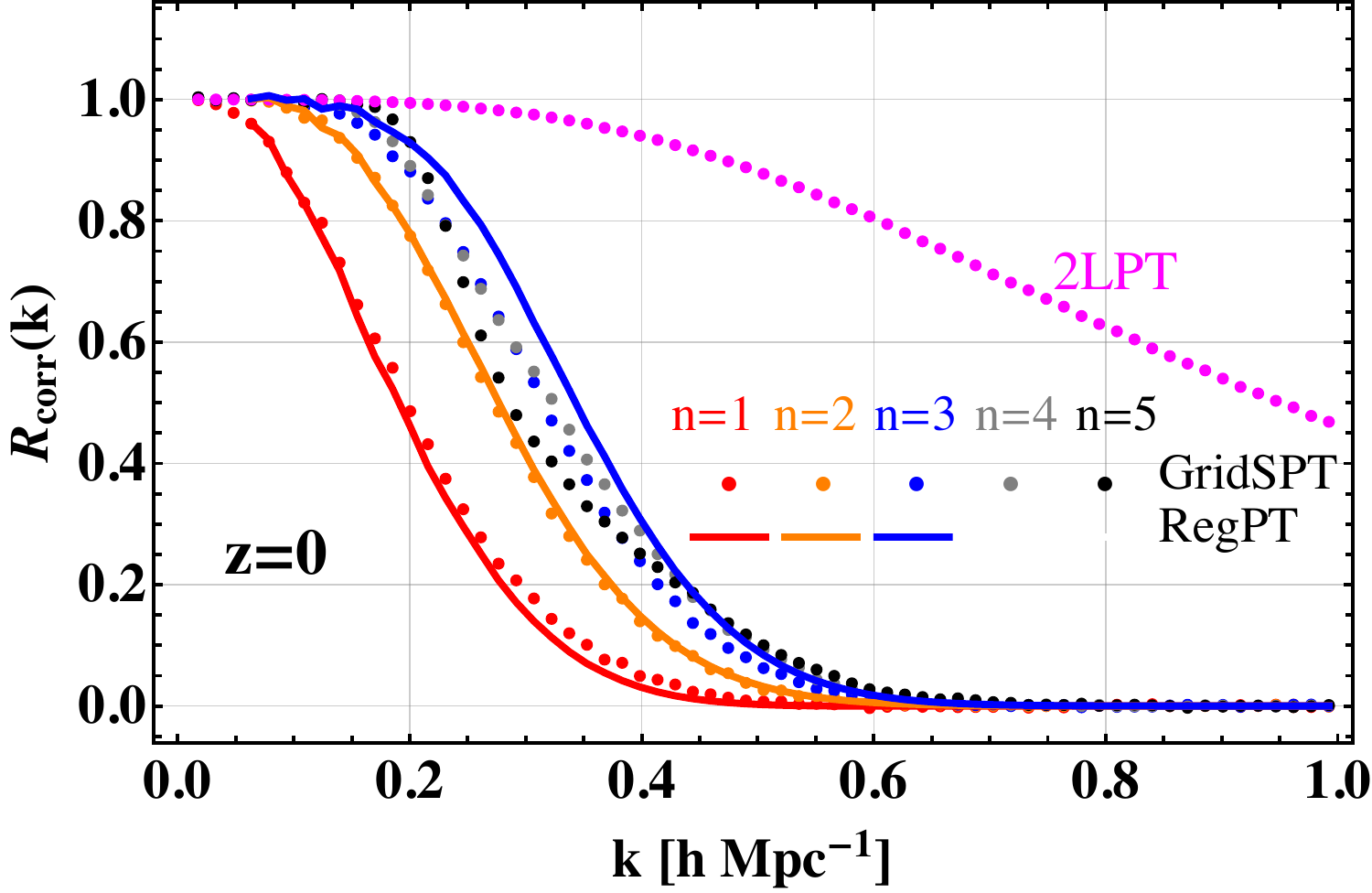}
\end{center}
\vspace*{-0.3cm}
\caption{Cross correlation coefficient for {\tt GridSPT} and $N$-body density fields, $R_{\rm cross}^{(n)}(k)$, defined at Eq.~(\ref{eq:r_corr_sum_nth}). The measured results at $z=1$ (left) and $0$ (right) are shown. Solid lines are the predictions based on the RegPT treatment, assuming that the evolved density field in $N$-body simulation is described with the multi-point propagator expansion at two-loop order  (see Appendix \ref{Appendix:pkcross_prediction}). For comparison, cross correlation coefficient for {\tt 2LPT} and $N$-body density fields is also shown in filled magenta circles. 
\label{fig:Cross_corr_pk_sum}
}
\end{figure*}

\subsubsection{Cross correlation}
\label{sec:pkcross}

In order to systematically compared the \texttt{GridSPT} density fields with $N$-body simulations, we calculate the cross-correlation between them. First, we consider the density field at each PT order, and compute the cross correlation {with the density field constructed from $N$-body simulations in Fourier space. Fig.~\ref{fig:Cross_corr_pk_each} shows the cross-correlation coefficients, $r_{\rm corr}(k)$, defined by
\begin{align}
r^{(n)}_{\rm corr}(k) \equiv\frac{\sum_{|\bfk|=k} \mbox{Re}\,\bigl[\delta_n(\bfk)\delta_{\rm N\mbox{-}body}(\bfk)\bigr] }{\sqrt{ \sum_{|\bfk|=k} |\delta_n(\bfk)|^2\,\sum_{|\bfk|=k} |\delta_{\rm N\mbox{-}body}(\bfk)|^2}}.  
\label{eq:r_corr_nthPT}
\end{align}
Note that $-1\leq r_{\rm corr}^{(n)} \leq1$. The measured results at $z=1$ (left) and $0$ (right) are shown up to the fifth order of \texttt{GridSPT} density fields. 

As shown in Fig.~\ref{fig:Cross_corr_pk_each}, all the correlations tend to get suppressed at high-$k$, and, at $z=0$, the suppression of correlation appears prominent even on large scales. This is indeed what is expected. An interesting point may be that the correlation of the higher-order PT fields $(n\geq2)$ with $N$-body is not monotonic, and exhibits anti-correlation ($r_{\rm corr}^{(n)}<0$) for a certain range of $k$. 

Indeed, these non-monotonic behaviors, together with a strong damping at high-$k$, are predicted by a resummed PT treatment. In Fig.~\ref{fig:Cross_corr_pk_each}, we also show (solid lines) the predictions of the same quantity based on the multi-point propagator expansion \cite{Bernardeau:2008fa} with regularized propagators, called RegPT \cite{Taruya:2012ut}. The RegPT predictions are made by assuming that the evolved density field in $N$-body simulation is described with the multi-point propagator expansion at two-loop order. In Appendix \ref{Appendix:pkcross_prediction}, we present a recipe to compute $r_{\rm cross}$, and derive the analytic expressions for the cross-power spectrum between SPT and nonlinear density fields, valid at the two-loop order. 

Apart from a small discrepancy in $n=3$, RegPT predictions agree well with measured results of cross-correlation coefficient. The discrepancy in the $n=3$ case is presumably due to the lack of higher-loop corrections, although the breakdown of the single-stream PT treatment might possibly play a role. In any case, an important consequence of this agreement is that the suppression of the measured correlation at high-$k$ comes from the randomness of the linear displacement field, which can be described with RegPT by a (partial) resummation of an infinite series of SPT expansion in the high-$k$ limit. This explains why the \texttt{GridSPT} density field with the {\it naive} SPT expansion looses quickly the correlations with the fully nonlinear density field at small scales. On the other hand, the non-monotonic behaviors at lower-$k$ is originated from the  SPT kernels, $F_n$, in Eq.~(\ref{eq:delta_n_Fourier}), which can can be both positive and negative. In other words, the non-monotonic behavior of $r_{\rm cross}^{(n)}$ directly manifests the poor convergence of the SPT expansion, and this explains why summing up each order of perturbation improves very slowly, as shown in Fig.~\ref{fig:Cross_corr_pk_sum}, where we plot the cross-correlation coefficient defined by
\begin{align}
R^{(n)}_{\rm corr}(k) \equiv\frac{\sum_{|\bfk|=k} \mbox{Re}\,\Bigl[\bigl(\sum_{i=1}^n D_+^n\,\delta_i(\bfk) \Bigr)\delta_{\rm N\mbox{-}body}(\bfk)\Bigr] }{\sqrt{ \sum_{|\bfk|=k} |\sum_{i=1}^n D_+^n\,\delta_i(\bfk)|^2\,\sum_{|\bfk|=k} |\delta_{\rm N\mbox{-}body}(\bfk)|^2}}.  
\label{eq:r_corr_sum_nth}
\end{align}

\begin{figure*}[tb]
\begin{center}
\includegraphics[width=8.7cm,angle=0]{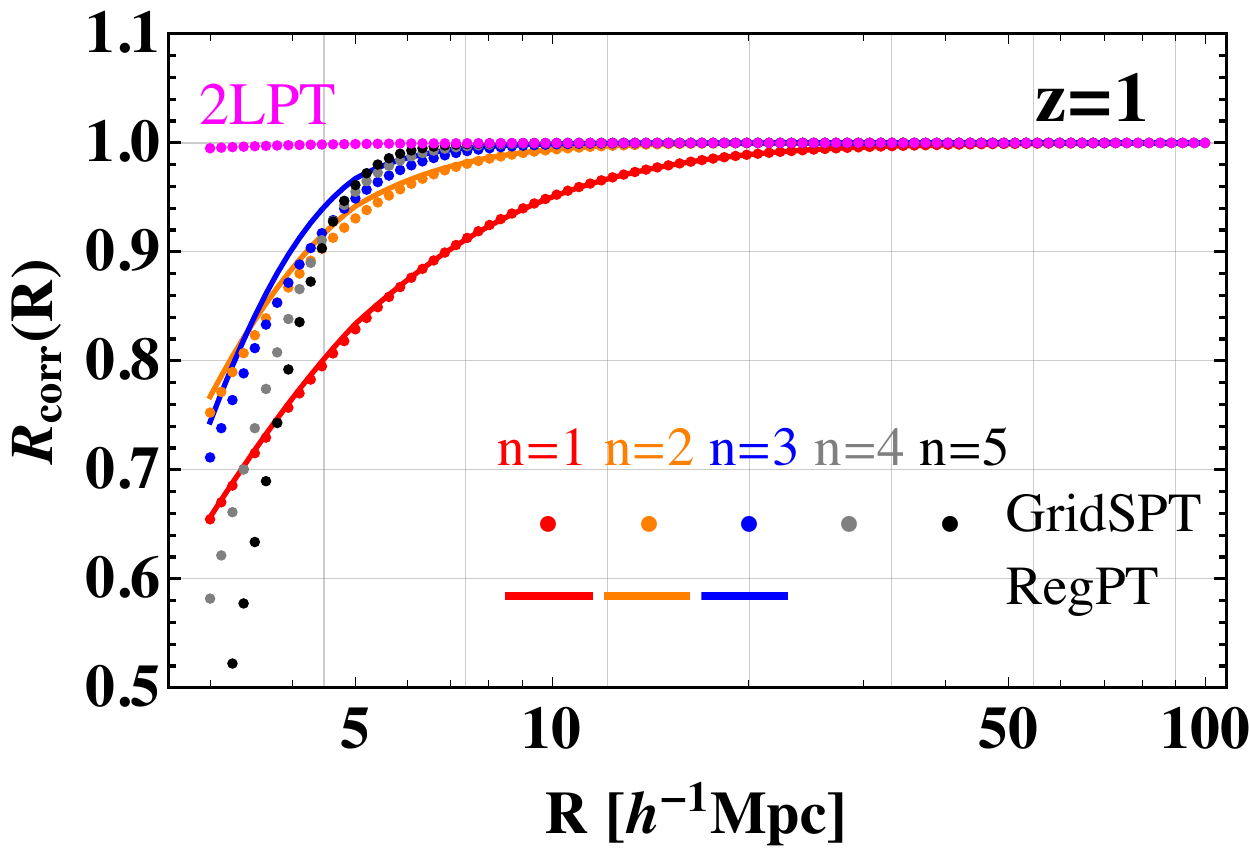}
\hspace*{0.2cm}
\includegraphics[width=8.7cm,angle=0]{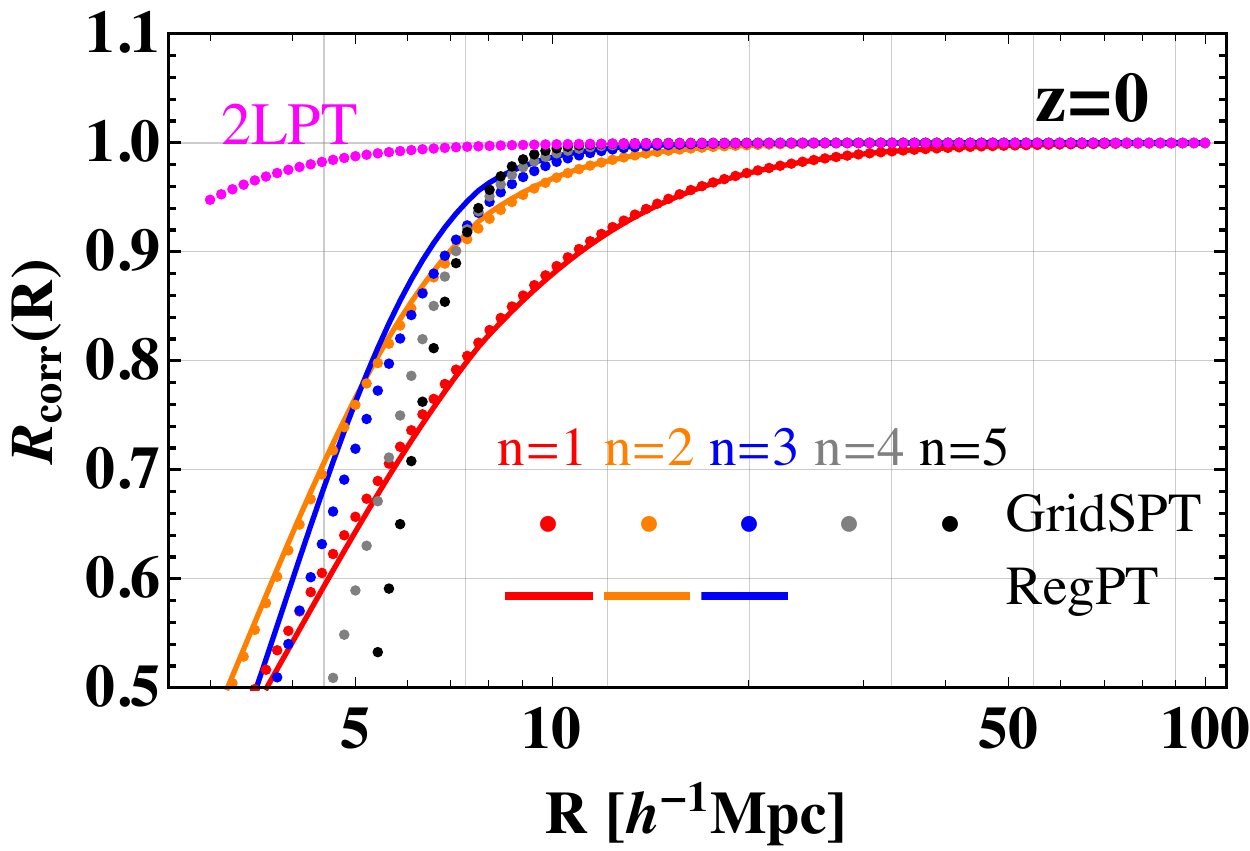}
\end{center}
\vspace*{-0.4cm}
\caption{Cross correlation coefficient for the {\tt GridSPT} and $N$-body density fields, 
$R^{(n)}_{\rm cross}(R)$, as similarly defined in Eq.~(\ref{eq:r_corr_sum_nth}) but 
with real-space density fields smoothing with Gaussian filter. The results are plotted as function of smoothing scale, $R$. Notation and line types are the same as in Fig.~\ref{fig:Cross_corr_pk_sum}. 
\label{fig:Cross_corr_moment}
}
\end{figure*}

As we can see from Fig.~\ref{fig:Cross_corr_pk_sum}, adding higher-order density fields recover the correlation at large scales to some extent, and the RegPT predictions, depicted as solid line, explain the measured results quantitatively up to $n=3$. At $n>3$, however, higher-order  corrections rather worsen the correlation with $N$-body simulation at intermediate scales around $k\sim0.5\,h$\,Mpc$^{-1}$. While this behavior might be possibly originated from small numerical flows as observed in the auto-power spectrum (see Fig.~\ref{fig:pkdd_1oop_2loop}), we see in Sec.~\ref{subsec:structure} that no severe cancellation of the higher-order terms has occurred at the density fields. Though the impact of numerical flaws inferred from Fig.~\ref{fig:pkdd_1oop_2loop} may result in  a $\sim10$\% systematic error, we rather suspect that substantial de-correlation found in $n>3$ cases is related to the UV-sensitive features of higher-order density fields seen in Fig.~\ref{fig:Slice_1D_Rdept}.

To clarify this point, we measure the cross-moments of the real-space density fields smoothed with a Gaussian filter. Varying the smoothing scales, we calculate the cross-correlation coefficients, which is defined similar to Eq.~(\ref{eq:r_corr_sum_nth}) in real-space, as a function of the smoothing scales $R$ in  Fig.~\ref{fig:Cross_corr_moment}. The scale-dependent behaviors seen in Fig.~\ref{fig:Cross_corr_moment} are what is anticipated from the Fourier-space cross correlation coefficients. That is, adding higher-order corrections does not improve the correlation at $n>3$, and rather leads to a de-correlation at small scales. The notable point is its characteristic scale, i.e., $R\sim 3-5\,h^{-1}$\,Mpc at $z=1$ and $R\sim 6-8\,h^{-1}$\,Mpc at $z=0$. The latter indeed matches the scales where we found a spurious wiggle feature in Fig.~\ref{fig:Slice_1D_Rdept}. We thus think that the behaviors seen in Figs.~\ref{fig:Cross_corr_pk_sum} and \ref{fig:Cross_corr_moment} are real, and capture the limitation of SPT. This point will be further discussed in detail in Sec.~\ref{sec:jointPDF}.

Finally, as a reference, we compare the results with those between second-order Lagrangian PT (2LPT) and $N$-body simulation, which are plotted in magenta filled circles in Figs.~\ref{fig:Cross_corr_pk_sum} and \ref{fig:Cross_corr_moment}. We find that the correlation obtained from {\tt 2LPT} is much better than those from SPT, indeed, at every redshift and on every scale. This is again due to the fact that the density fields generated with Lagrangian PT is constructed based on the displacement
of the particles which respects the mass conservation (that is, Lagrangian PT ensures that $\delta\ge-1$). It can therefore describe the large-scale matter flow, at a certain precision, in the single-stream regime. In contrast, the SPT requires a (partial) re-summation of an infinite series of PT expansion to describe such an effect. Nevertheless, as shown in Figs.~\ref{fig:pkdd_1oop_2loop}-\ref{fig:bkdd_50realizations}, a better performance in the cross-correlation coefficients does not necessarily imply that the Lagrangian PT always gives a better statistical prediction for the mass distribution. Indeed, the density fields generated with {\tt 2LPT} generally under-predicts the amplitude of peaks (see Fig.~\ref{fig:Slice_1D_high}), and this can result in the underestimation of the power spectrum and bispectrum amplitudes. In this respect, SPT suits better for the statistical predictions sensitive to the high-density regions. In other words, at high-density regions, the statistical correlation of SPT with $N$-body simulation may be better than that of {\tt 2LPT}. We shall show it below by the direct measurement of the correlation in the point-by-point manner.

\begin{figure*}[tb]
\begin{center}
\includegraphics[width=16cm,angle=0]{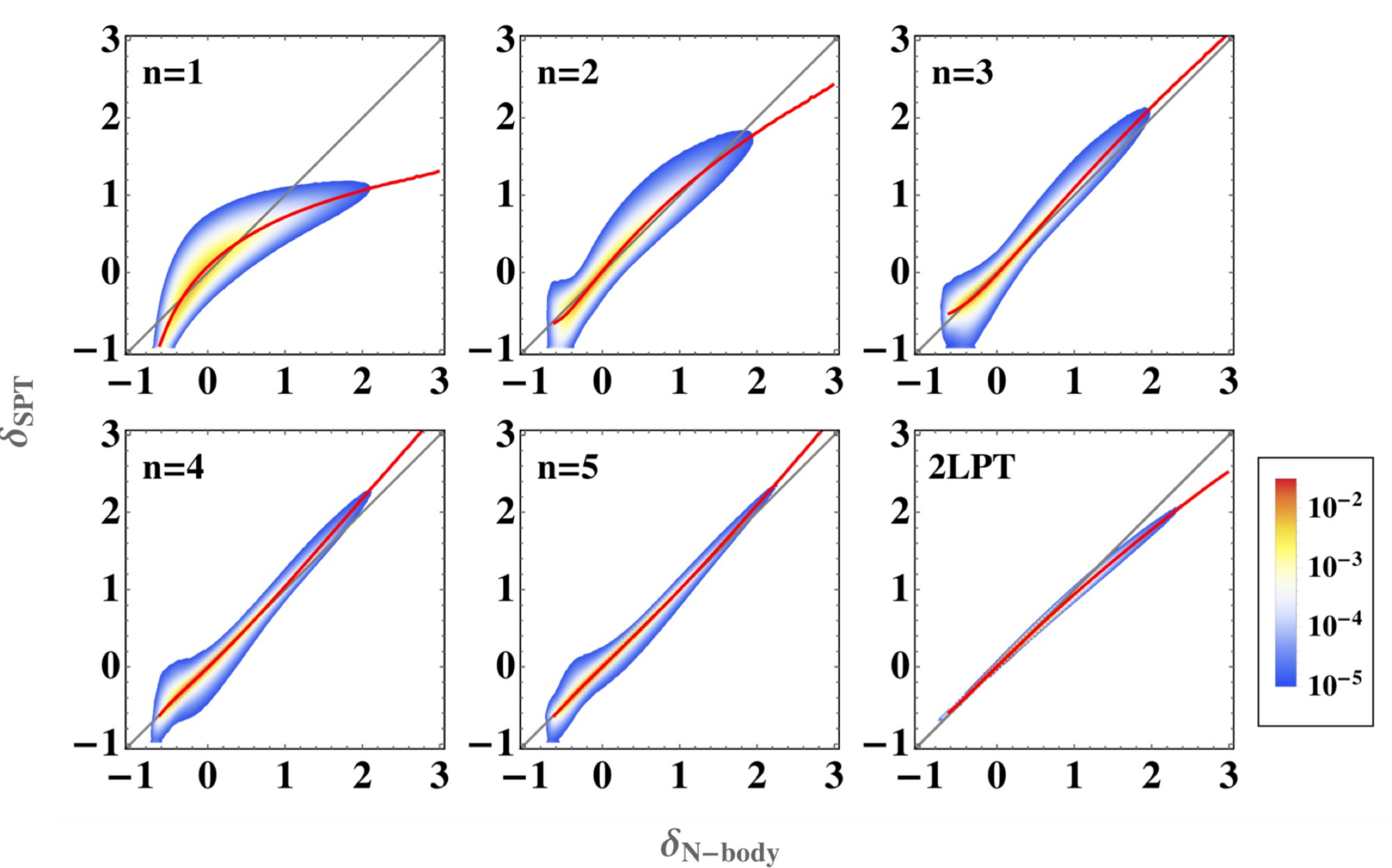}
\end{center}

\vspace*{-0.5cm}

\caption{Joint PDF of the {\tt GridSPT} and $N$-body density fields, $\Prob(\delta_{\rm SPT},\delta_{\rm N\mbox{-}body})$, where $\delta_{\rm SPT}=\sum_{i=1}^n\,D_+^i\,\delta_i$ with the number $n$ indicated in each panel. In measuring joint PDF, the Gaussian filter of $R=10\,h^{-1}$Mpc is applied to the final density field, and the results at $z=0$ are shown. For comparison, we also plot in bottom right panel the joint PDF of the second-order Lagrangian PT and $N$-body density fields. 
\label{fig:jointPDF_R10}}
\end{figure*}

\begin{figure*}[tb]
\begin{center}

\includegraphics[width=15.2cm,angle=0]{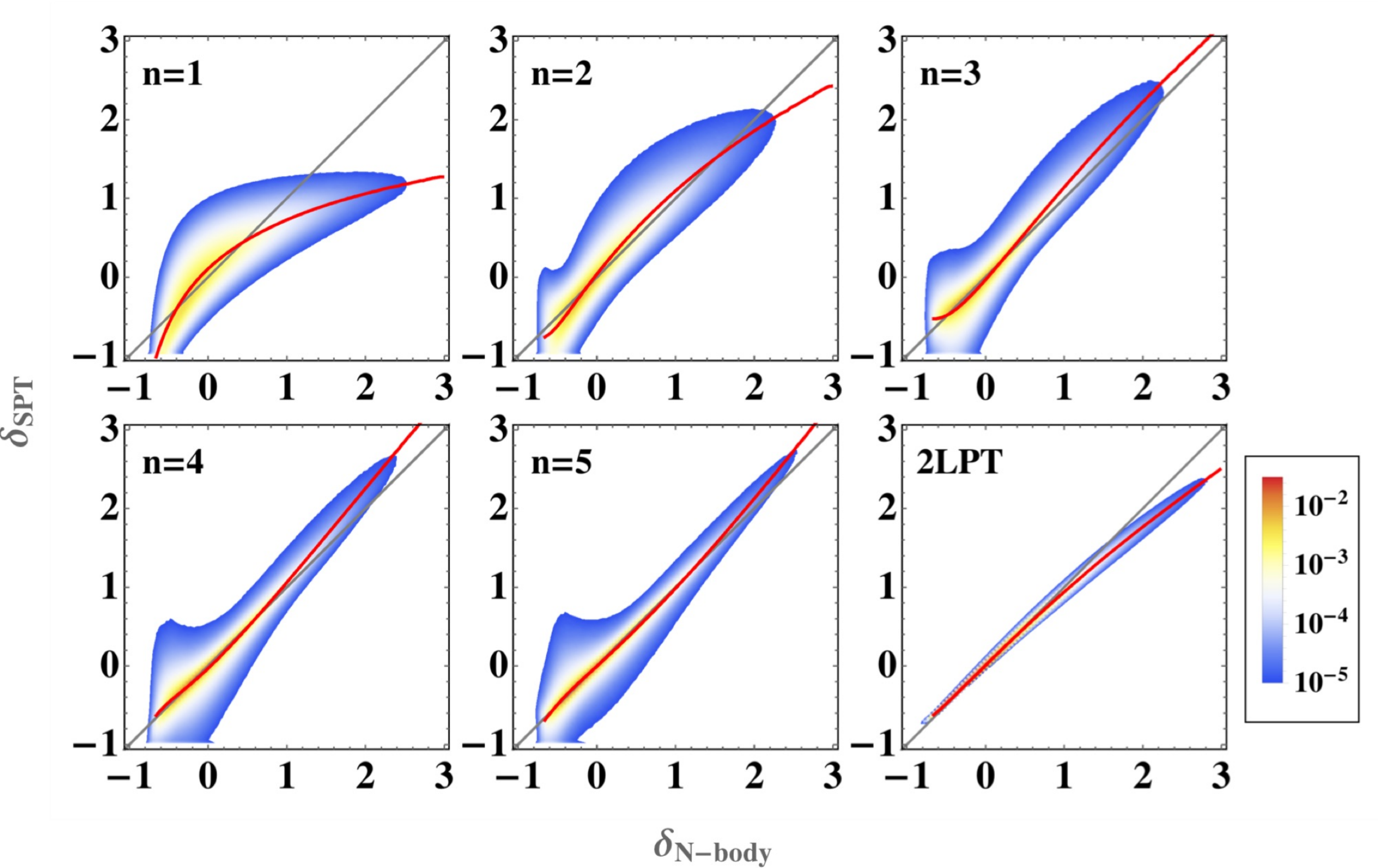}


\includegraphics[width=15.2cm,angle=0]{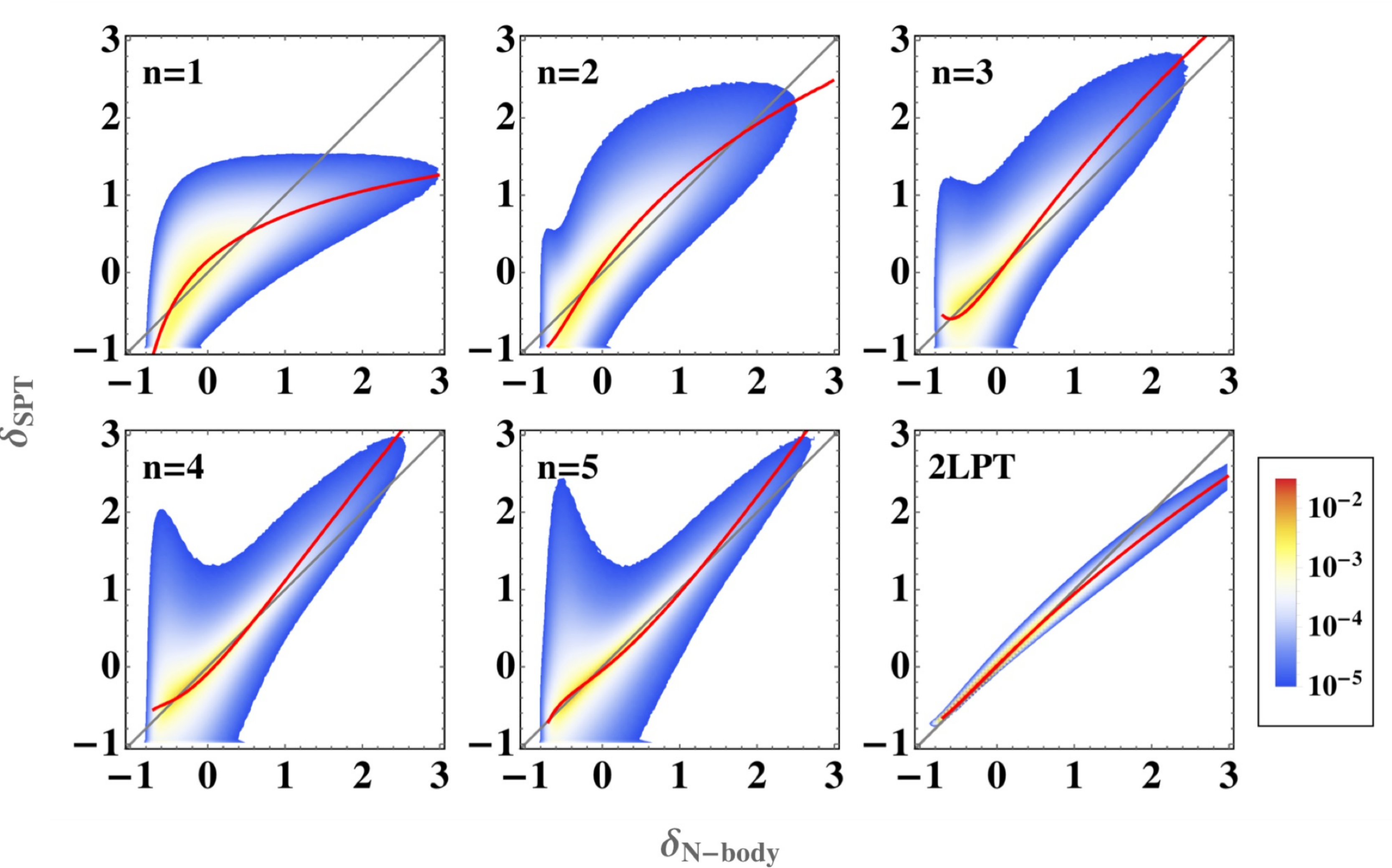}
\end{center}
\vspace*{-0.5cm}

\caption{Same as in Fig.~\ref{fig:jointPDF_R10}, but the results with different smoothing scales applied are shown: $R=8$ (top) and $6\,h^{-1}$Mpc (bottom).
\label{fig:jointPDF_R8_6}}
\end{figure*}

\subsubsection{Joint probability distribution of GridSPT and N-body simulation}
\label{sec:jointPDF}

Our last measurement is the joint probability distribution function (PDF) for the local density contrast obtained both from the \texttt{GridSPT} calculation, $\delta_{\rm SPT}=\sum_{i=1}^n\,D_+^i \delta_i$, and $N$-body simulation, $\delta_{\rm N\mbox{-}body}$. This quantity, denoted by $\Prob(\delta_{\rm SPT},\delta_{\rm N\mbox{-}body})$, characterizes the probability that at a given position in the space, the overdensity from $N$-body simulation is $\delta_{\rm N\mbox{-}body}$, and the \texttt{GridSPT} calculation predicts $\delta_{\rm SPT}$. It thus gives the point-by-point comparison of the two density fields.

Fig.~\ref{fig:jointPDF_R10} shows the measured result of the joint PDF for smoothed density fields with Gaussian filter at $z=0$. Here, we show the results with the smoothing radius $R=10\,h^{-1}$Mpc. In each panel, the results with GridSPT are presented up to the fifth order (i.e., $n=1-5$). For reference, the joint PDF between {\tt 2LPT} and $N$-body density fields is also shown (lower right). Compared to the results with {\tt 2LPT}, the lower-order results of the \texttt{GridSPT} calculations show not only a large scatter but also a declined correlation feature in the joint PDF. Here, the magenta line in each panel represents the conditional mean for a given $N$-body density field, $\overline\delta_{\rm SPT}(\delta_{\rm N\mbox{-}body})$, defined by:
\begin{align}
\overline\delta_{\rm SPT}(\delta_{\rm N\mbox{-}body}) = \int d \delta_{\rm SPT}\,\delta_{\rm SPT}\,\frac{\Prob(\delta_{\rm SPT},\,\delta_{\rm N\mbox{-}body})}{\Prob(\delta_{\rm N\mbox{-}body})}.
\end{align}

As increasing $n$ (i.e., the order of perturbation), the correlation gets tighter, and the conditional mean tends to follow a linear relation depicted as black solid line, meaning that the \texttt{GridSPT} density field gets closer to the $N$-body density field. Overall, the result with {\tt 2LPT} still looks much better as the correlation is much tighter and closer to the linear relation. However, a closer look at the high-density region reveals a small tilt i.e., $\overline{\delta}_{\rm 2LPT}(\delta_{\rm N\mbox{-}body})\,<\,\delta_{\rm N\mbox{-}body}$, where the \texttt{GridSPT} results at $n>3$ are apparently better correlated with the $N$-body results. Beyond this regime, on the other hand, the conditional mean of the \texttt{GridSPT} becomes steeper, and thus the \texttt{GridSPT} tends to over-predicts the amplitude of density fields. These behaviors are indeed what we saw in the power spectrum and bispectrum (Figs.~\ref{fig:pkdd_1oop_2loop}-\ref{fig:bkdd_50realizations}), and partly explains why SPT gives a better prediction.

Finally, as we discussed in Sec.~\ref{subsec:structure}, the measured result of joint PDF is also sensitive to the choice of filter scales. Fig.~\ref{fig:jointPDF_R8_6} shows the same plot as in Fig.~\ref{fig:jointPDF_R10}, but the results of the Gaussian filter of the radii $R=8$ (top) and $6\,h^{-1}$ Mpc (bottom) are shown. Decreasing the filter scale, a sizable amount of scatter appears in the joint PDF of the \texttt{GridSPT} calculation at higher-order. In particular, the scatter becomes developed at the low-density region, and it extends to the un-physical region, $\delta<-1$. This is another manifestation of what we see in Fig.~\ref{fig:Slice_1D_Rdept}, This is presumably originated from the UV-sensitive mode-coupling behavior in the SPT, which explains why the correlation between \texttt{GridSPT} and $N$-body results becomes substantially reduced at higher-order, as seen in Figs.~\ref{fig:Cross_corr_pk_sum} and \ref{fig:Cross_corr_moment}.

\section{Conclusion}
\label{sec:conclusion}

In this paper, we have presented a novel, grid-based calculation for standard perturbation theory (SPT) of large-scale structure, with which a face-to-face comparison of the PT prediction with $N$-body simulations is made possible. With the FFT, the code, {\tt GridSPT}, written in {\tt c++},  can quickly compute the higher-order density fields based on the real-space recursion formula in Eq.~(\ref{eq:recursion_formula}). Then, we have demonstrated the grid-based PT calculation up to the fifth order, and studied the morphological and statistical properties of the SPT density fields in comparison with $N$-body simulations and Lagrangian PT predictions. 

Our findings are summarized as follows:

\begin{itemize}
 \item Increasing the PT order up to more than third order, the SPT reproduces the structure of density fields in $N$-body simulation reasonably well on large scales ($R\geq10\,h^{-1}$\,Mpc). In particular, the convergence of the PT expansion is found to be faster at higher-density region, while it conversely gets slower at lower-density regions.
 \item On the other hand, on small scales, the SPT is prone to produce spurious and un-physical structure in the density field. The statistical correlation with $N$-body simulations is generally poor on small scales ($k\gtrsim0.2\,h$\,Mpc$^{-1}$ or $R\lesssim10\,h^{-1}$Mpc at $z=0$), and including the higher-order corrections does not improve the cross correlation at all. 
 \item In contrast, the second-order Lagrangian PT (2LPT) prediction gives a better correlation with $N$-body density field even at the second order. In particular, it reproduces the structure at low-density regions remarkably well. However, the Lagrangian PT systematically underestimates the amplitude at higher-density regions, and this can lead to a poor statistical description of the power spectrum and bispectrum. Rather, the SPT gives a better statistical prediction, and albeit the lack of precision, it can qualitatively capture the trend of nonlinear growth.
\end{itemize}

The deficiencies of the SPT predictions certainly come from the facts that the present expansion scheme does not respect the mass conservation at finite order (that is, regions with $\delta<-1$ exist), and the single-stream approximation in the PT treatment is invalid on small scales. Our findings are thus regarded as a direct manifestation of these deficiencies both at field and statistical levels. To remedy these drawbacks, resummed PT scheme helps the convergence of PT expansion (e.g., \cite{Crocce:2005xy,Crocce:2007dt,Bernardeau:2008fa,Taruya:2007xy,Pietroni:2008jx,Taruya:2009ir,Crocce:2012fa,Taruya:2012ut}), and can mitigate the impact of violating the mass conservation. Further, the effective-field-theory approach (e.g., \cite{2012JCAP...07..051B,2012JHEP...09..082C,2014PhRvD..89d3521H,Baldauf:2015aha}) would be crucial to remedy the UV sensitive behavior in SPT arising from the violation of single-stream flows. Although various works have been so far done, most of the works has been made at the level of statistical quantities, not at the field level. Development of resummation scheme and effective-field-theory treatment at the field level is thus rather interesting subject.

Finally, although the present \texttt{GridSPT} treatment is still primitive and needs to be improved for the practical purposes along this line, possible application of {\texttt{GridSPT} calculation may be worthy of consideration. One interesting application is obviously a direct comparison with observed galaxy distributions, after implementing appropriate prescriptions for galaxy bias and redshift-space distortions. There has been recently a general description of the galaxy bias exploited on the basis of the SPT \cite{Desjacques_Jeong_Schmidt_review2018,Mirbabayi_etal2015,Desjacques_Jeong_Schmidt2018}, and thus the implementation of galaxy bias is straightforward at the field level. Also, the redshift-space distortions are described by a simple mapping formula, and thus easy to handle at the field level. Taking one step further, the method may be applied to the reconstruction problem to infer the initial conditions of the universe (e.g., \cite{Jasche_Wandelt2013,Wang_etal2013,Kitaura2013,Wang_etal2014,Schmittfull_etal2017,Seljak_etal2017,Shi_Cautun_Li2018,Modi_Feng_Seljak2018} for recent works). Since the generation of higher-order density fields is rather fast, with the help of a sophisticated algorithm for optimization, the grid-based PT approach may have a potential to efficiently extract the cosmological information at weakly nonlinear regime, rather than the statistical PT calculation.

\acknowledgements{
This work was supported in part by MEXT/JSPS KAKENHI Grant Number JP15H05899 and JP16H03977 (AT), and JP17K14273 (TN). TN also acknowledges financial support from Japan Science and Technology Agency (JST) CREST Grant Number JPMJCR1414. DJ acknowledges support from NSF grant (AST-1517363) and NASA 80NSSC18K1103. 
Numerical computation was partly carried out at the Yukawa Institute Computer Facility.
}

\appendix
\section{Sensitivity of {\tt GridSPT} calculations to high-$k$ cutoff} 
\label{Appendix:cutoff_depdendence}

In this Appendix, we investigate the impact of the high-$k$ cutoff for the sharp-$k$ filter on the {\tt GridSPT} calculations, particularly paying an attention to the systematic error associated with the aliasing effect. 

As we discussed in Sec.~\ref{sec:gridSPT}, high-$k$ modes of $|k_{x,y,z}|>k_{\rm crit}=(2\pi/L_{\rm box})(N_{\rm grid}^{1/3}/3)\simeq1.07\,h$\,Mpc$^{-1}$ can produce the non-vanishing spurious contribution (aliasing error) to the grid-based PT calculations through the non-linear interaction. These modes are to be subtracted, and in this paper, we adopt the isotropic sharp-k filter, with which modes with $|\bfk|>k_{\rm cut}$ are set to zero at each order of PT calculation. In general, a smaller value of the cutoff wavenumber is preferred for a secure removal of the aliasing error, but this would change the mode-coupling behavior, also affecting the power spectrum and bispectrum predictions. It is thus important to find an optimal cutoff scale.

\begin{figure*}[tb]
\begin{center}
\includegraphics[width=8.7cm,angle=0]{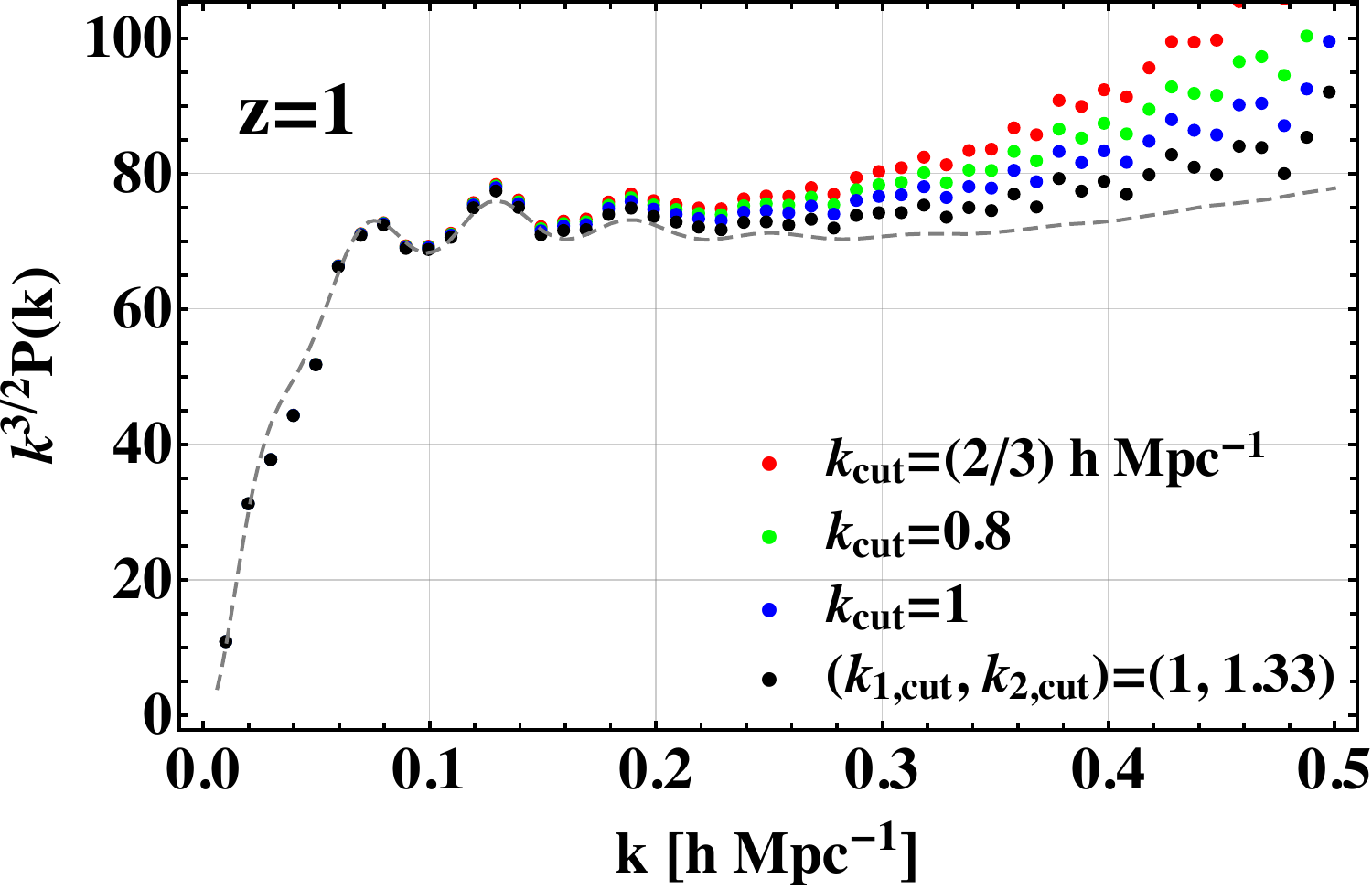}
\hspace*{0.2cm}
\includegraphics[width=8.7cm,angle=0]{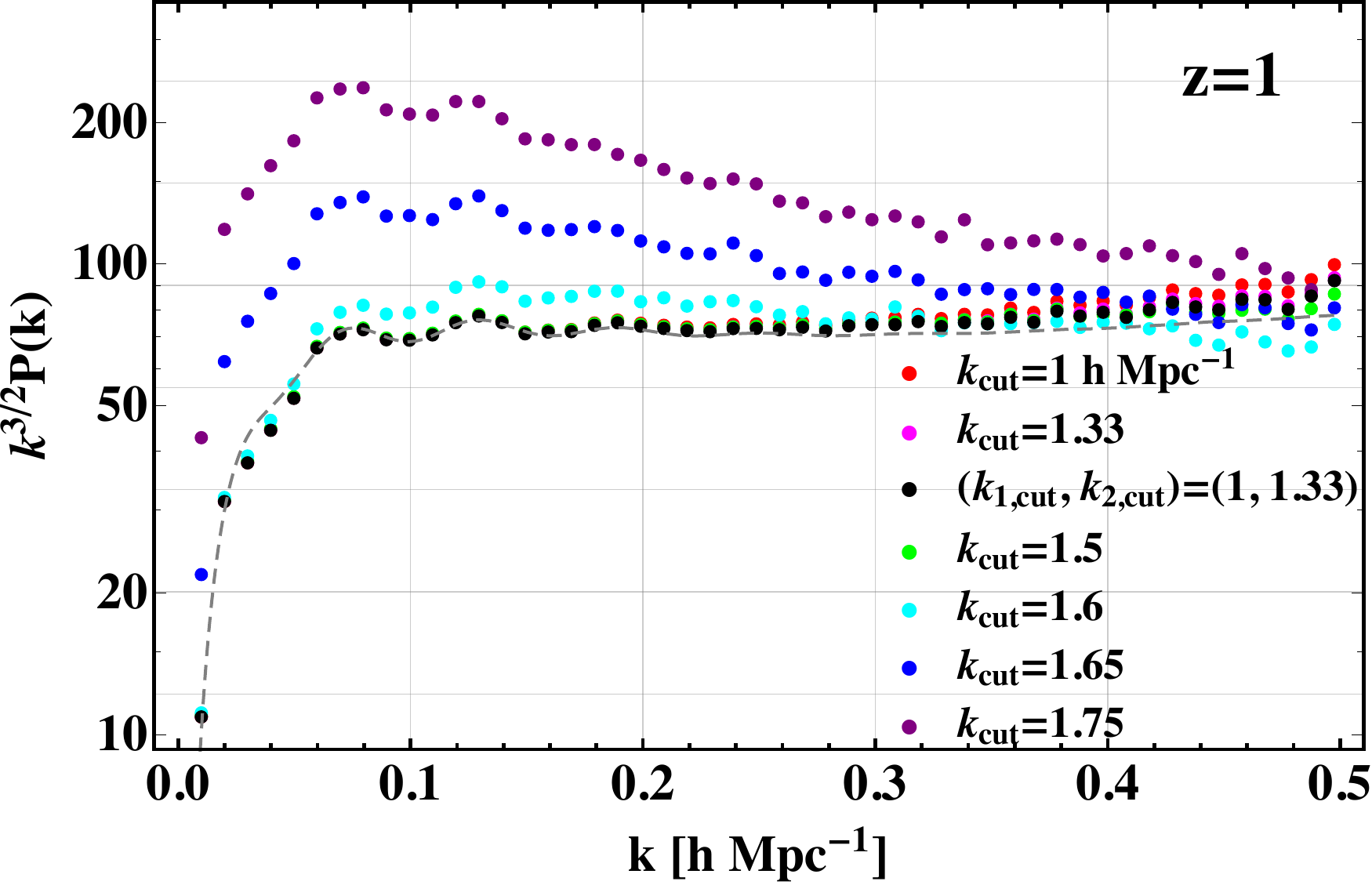}
\end{center}
\vspace*{-0.3cm}
\caption{Sensitivity of the two-loop power spectra to the high-$k$ cutoff in {\tt GridSPT} calculations. Varying $k_{\rm cut}$ of sharp-$k$ filter, the results  at $z=1$ are particularly shown. Left panel plots the cases with $k_{\rm cut}\lesssim k_{\rm crit}$, while right panel shows the results with $k_{\rm cut}\gtrsim k_{\rm crit}$, whose amplitudes are plotted in logarithmic scales. In both panels, black filled circles represent the results with the cutoff $k_{\rm crit,1}=1\,h$\,Mpc$^{-1}$ for the linear density field and $k_{\rm crit,2}=(4/3)\,h$\,Mpc$^{-1}$ for the higher-order PT density fields, which are the default setup used in Sec.~\ref{sec:demonstration}. For reference, gray dashed lines are the analytic PT predictions adopting the Nyquist frequency determined by the inter-particle distance of the $N$-body simulation as the high-$k$ cutoff. 
\label{fig:pk_gridSPT_cutoff}
}
\vspace*{0.2cm}
\begin{center}
\includegraphics[width=8.7cm,angle=0]{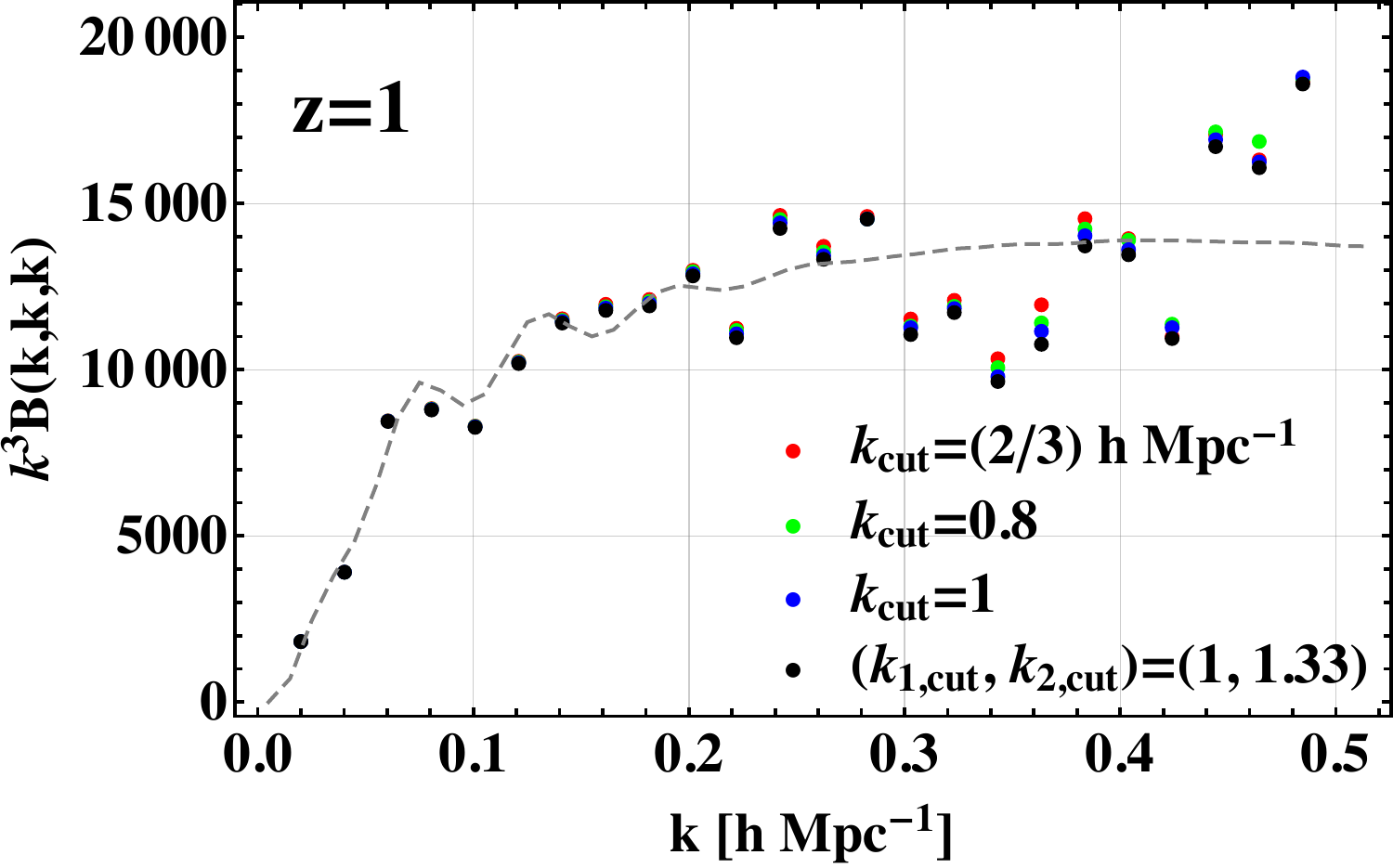}
\hspace*{0.2cm}
\includegraphics[width=8.7cm,angle=0]{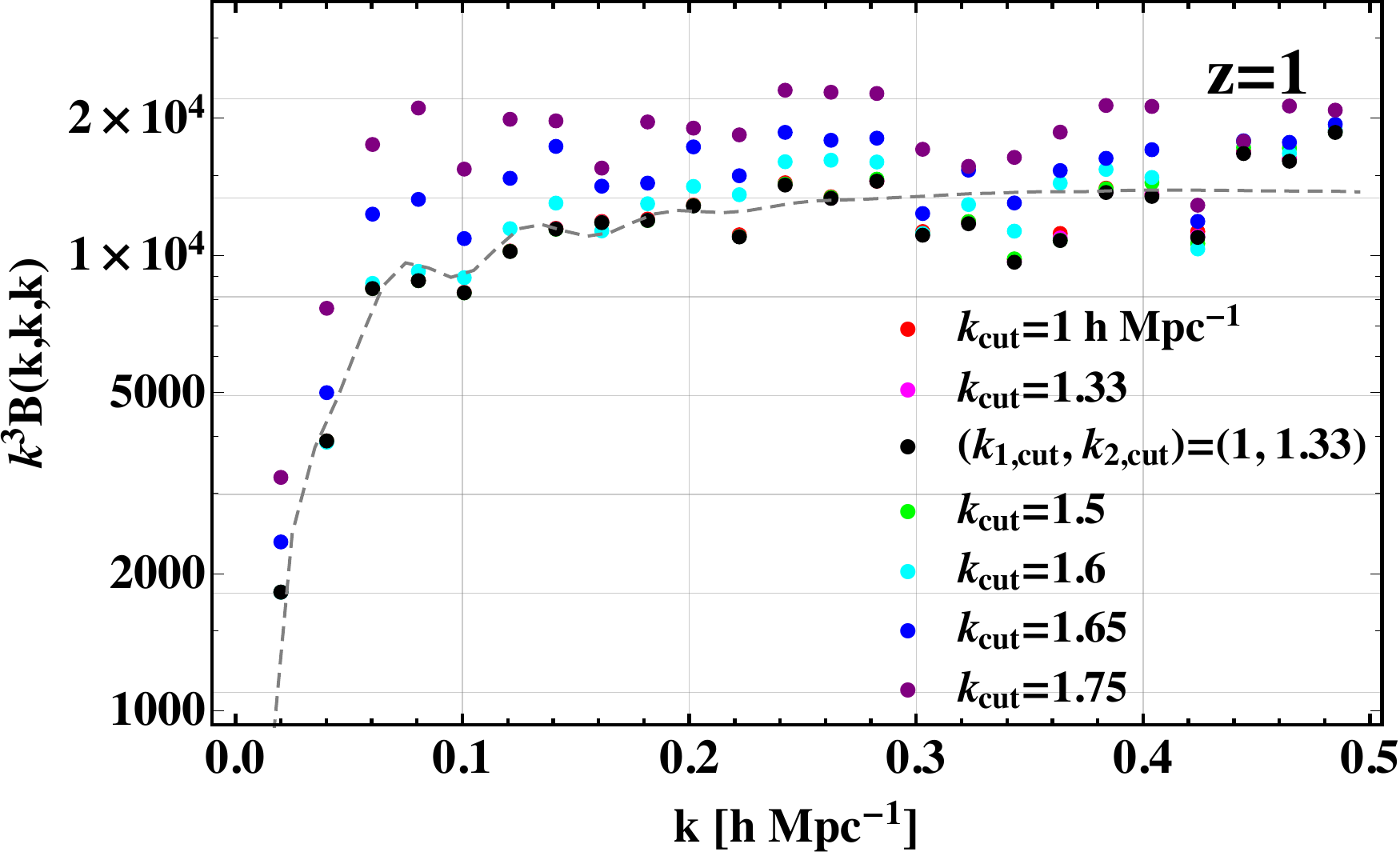}
\end{center}
\vspace*{-0.3cm}
\caption{Sensitivity of the one-loop bispectra to the high-$k$ cutoff in {\tt GridSPT} calculations. Left and right panel respectively show the results with $k_{\rm cut}\lesssim k_{\rm crit}$ and $k_{\rm cut}\gtrsim k_{\rm crit}$, with the latter case particularly plotted in logarithmic scales. Meanings of the symbols and lines are the same as in Fig.~\ref{fig:pk_gridSPT_cutoff}. 
\label{fig:bk_gridSPT_cutoff}
}
\end{figure*}

To see how the choice of cutoff scale affects the {\tt GridSPT} results, we plot in Figs.~\ref{fig:pk_gridSPT_cutoff} and \ref{fig:bk_gridSPT_cutoff} respectively the power spectrum prediction at two-loop order and bispectrum prediction at one-loop order at $z=1$, varying the cutoff scales. In each figure, left panels examine the cases with $k_{\rm cut}\lesssim k_{\rm crit}$, while right panels show the results with $k_{\rm cut}\gtrsim k_{\rm crit}$, whose amplitudes are plotted in logarithmic scales. For references, gray dashed lines are the analytic PT predictions with the high-$k$ cutoff $k_{\rm cut}=\pi/(L_{\rm box}/N_{\rm particle}^{1/3})=3.217\,h$\,Mpc$^{-1}$, corresponding to the Nyquist frequency determined by the inter-particle distance in our $N$-body simulations with $L_{\rm box}=1,000\,h^{-1}\,$Mpc and $N_{\rm particle}=1,024^3$ (that is, these predictions slightly differ from those shown in Figs.~\ref{fig:pkdd_1oop_2loop}-\ref{fig:bkdd_50realizations}, but the differences are subtle). 

As we see, a small $k_{\rm cut}$ enhances the small-scale amplitudes of the power spectrum and bispectrum. This is expected from the fact that the mode transfer generally occurs from large to small scales in the CDM spectrum, and introducing the high-$k$ cutoff makes the mode transfer at smaller scales ineffective compared to that at larger scales (see, e.q., Ref.~\cite{Nishimichi:2014rra}). On the other hand, the results with larger $k_{\rm cut}$ exhibit a strong enhancement in amplitudes at large scales, and a slight change of the cutoff scale largely alters the low-$k$ behaviors in power spectrum and bispectrum. While this behavior is expected and is basically explained by the spurious aliasing contributions, a notable point is that the effect appears prominent at $k_{\rm cut}\gtrsim1.6\,h$\,Mpc$^{-1}$, which is larger than the critical wavenumber $k_{\rm crit}\simeq1.07\,h$\,Mpc$^{-1}$.

A possible reason for this may be that we use the isotropic filter, and even if the cutoff wavenumber slightly exceeds $k_{\rm crit}$, majority of the high-$k$ modes that can produce the aliasing error is set to zero except the modes close to the $k_{x,y,z}$ axes. Although the isotropic sharp-$k$ filter ceases to be effective at $k_{\rm cut}>\sqrt{3}\,k_{\rm crit}$, a part of the aliasing error is still suppressed if the cutoff scale is chosen below $\sqrt{3}\,k_{\rm crit}$. Based on this consideration, one may adopt the cutoff scale of the isotropic filter larger than $k_{\rm crit}$. Indeed, we find empirically that applying the sharp-$k$ filter with $k_{\rm crit,1}=1\,h$\,Mpc$^{-1}$ for the linear density field and $k_{\rm crit,2}=(4/3)\,h$\,Mpc$^{-1}$ for the PT density fields higher than second order, the {\tt GridSPT} calculations give the best performance to the power spectrum and bispectrum predictions amongst those we examined, depicted as filled black symbols in Figs.~\ref{fig:pk_gridSPT_cutoff} and \ref{fig:bk_gridSPT_cutoff}. We have also compared other quantities such as cross correlation coefficients, $r_{\rm corr}$ and $R_{\rm corr}$, and joint PDF with those obtained with the cutoff scale $k_{\rm cut}<k_{\rm crit}$ , and checked that all the results are hardly changed. Hence, we decided to adopt this choice as default setup and presented the {\tt GridSPT} results.

\section{Cross-correlation coefficient from RegPT}
\label{Appendix:pkcross_prediction}

In this Appendix, based on the resummed PT calculation with RegPT, we describe a prescription to compute $r_{\rm corr}(k)$ and $R_{\rm corr}(k)$ shown in solid lines of Figs.~\ref{fig:Cross_corr_pk_each} and \ref{fig:Cross_corr_pk_sum} is presented.

Let us first rewrite Eqs.~(\ref{eq:r_corr_nthPT}) and (\ref{eq:r_corr_sum_nth}) with those in the continuum limit: 
\begin{align}
& r^{(n)}_{\rm corr}(k)=\frac{\PkCross^{(n)}(k)}{\PkSPT^{(nn)}(k) P_{\rm N\mbox{-}body}(k)}, 
\label{eq:rcorr_continuum}
\\
& R^{(n)}_{\rm corr}(k)=\frac{\sum_{i=1}^n \PkCross^{(i)}(k)}{\sum_{i\leq j}^n \PkSPT^{(ij)}(k) P_{\rm N\mbox{-}body}(k)}. 
\label{eq:Rcorr_continuum}
\end{align}
In the above, $P_{\rm N\mbox{-}body}$ is the power spectrum of the evolved density field in $N$-body simulations, for which we use the measured data. On the other hand, the quantities $\PkCross^{(n)}$ and $\PkSPT^{(ij)}$, which involve the PT density fields, are defined by:
\begin{align}
& \langle\delta_n(\bfk)\,\delta_{\rm N\mbox{-}body}(\bfk')\rangle = (2\pi)^3\,\delta_{\rm D}(\bfk+\bfk')\,\PkCross^{(n)}(k),
\label{eq:def_pkcross}
\\
& \langle\delta_i(\bfk)\,\delta_j(\bfk')\rangle = (2\pi)^3\,\delta_{\rm D}(\bfk+\bfk')\,\PkSPT^{(ij)}(k),
\label{eq:def_pkSPTnn}
\end{align}
These are what we want to deal with based on the analytic PT treatment. Below, we present their analytic expressions up to $n=3$.

The quantity, $\PkSPT^{(ij)}$, appears frequently in the statistical calculation of power spectrum in SPT.  With the expression given at Eq.~(\ref{eq:delta_n_Fourier}), the Gaussianity of the initial density field $\delta_0$ leads to the following non-vanishing expressions relevant up to $n=3$ (e.g., Refs.~\cite{Fry:1993bj,Carlson:2009it,Taruya:2009ir}):  
\begin{align}
 \PkSPT^{(11)}(k)&= P_0(k)
\label{eq:pkSPT11}
\\
 \PkSPT^{(22)}(k)&= 2\, \int  \frac{d^3\bfp }{(2\pi)^3} \{ F_2(\bfk, \bfk-\bfp)\}^2\,P_0(p)P_0(|\bfk-\bfp|),
\label{eq:pkSPT22}
\\
 \PkSPT^{(13)}(k)&= 6\,P_0(k) \int  \frac{d^3\bfp }{(2\pi)^3} F_3(\bfk, \bfp,-\bfp)\,2 P_0(p),
\label{eq:pkSPT13}
\\
 \PkSPT^{(33)}(k)&= 6 \,\int \frac{d^3\bfp d^3\bfq}{(2\pi)^6} \{F_3(\bfp,\bfq,\bfk-\bfp-\bfq) \,\}^2 P_0(p) P_0(q) P_0(|\bfk-\bfp-\bfq|) 
 + 9\, P_0(k) \Bigl[\int \frac{d^3\bfp}{(2\pi)^3}\,F_3(\bfk,\bfp,-\bfp) P_0(p)\Bigr]^2.
\label{eq:pkSPT33}
\end{align}
On the other hand, to derive the analytic expression of $\PkCross^{(n)}$, we need several steps. First, we use Eq.~(\ref{eq:delta_n_Fourier}) to express
\begin{align}
 \langle\delta_n(\bfk)\delta_{\rm N\mbox{-}body}(\bfk')\rangle
& = \int \frac{d^3\bfp_1\cdots d^3\bfp_n}{(2\pi)^{3(n-1)}}\,F_n(\bfp_1,\cdots,\bfp_n)\,
\Bigl\langle\delta_0(\bfp_1)\cdots\delta_0(\bfp_n)\,\delta_{\rm N\mbox{-}body}(\bfk')\Bigr\rangle. 
\label{eq:cross_pk_LHS}
\end{align}
To rewrite the ensemble average at right-hand side, we identify $\delta_{\rm N\mbox{-}body}$ with the nonlinear density field which can be dealt with single-stream treatment based on Eqs.~(\ref{eq:eq_continuity})-(\ref{eq:eq_Poisson}). While this is the assumption valid in the single-stream regime, the density field is not necessarily treated as small quantity.  We then introduce the multi-point propagator \cite{Bernardeau:2008fa}. The $(p+1)$-point propagator, denoted by $\Gamma^{(p)}$, is defined as
\begin{align}
&\frac{1}{p!}\left\langle
\frac{\delta^p\delta_{\rm N\mbox{-}body}(\bfk,\eta)}{\delta\delta_0(\bfk_1)
\cdots \delta\delta_0(\bfk_p)}\right\rangle =\delta_{\rm D}
(\bfk-\bfk_{1\cdots p})\,
\frac{1}{(2\pi)^{3(p-1)}} \,
\Gamma^{(p)}(\bfk_1,\cdots,\bfk_p).  
\end{align}
In the Gaussian initial condition relevant for our setup, the above definition is reduced to the following expression (e.g., \cite{Crocce:2005xz,Bernardeau:2008fa}):
\begin{align}
\Bigl\langle\delta_0(\bfk_1)\cdots\delta_0(\bfk_p)\delta_{\rm N\mbox{-}body}(\bfk)\Bigr\rangle_c=
n!\,(2\pi)^3\,\delta_{\rm D}(\bfk+\bfk_{1\cdots p})\,\Gamma^{(p)}(\bfk_1,\cdots,\bfk_p)\,P_0(k_1)\cdots P_0(k_p), 
\end{align}
where $\langle\cdots\rangle_c$ stands for the cumulant. Using the above, the moment correlation, $\langle\delta_0(\bfp_1)\cdots\delta_0(\bfp)\delta_{\rm N\mbox{-}body}(\bfk')\rangle$, is expressed as follows:
\begin{align}
& \Bigl\langle\delta_0(\bfp)\delta_{\rm N\mbox{-}body}(\bfk')\Bigr\rangle = 
(2\pi)^3\,\delta_{\rm D}(\bfp+\bfk')\,\Gamma^{(1)}(k')\,P_0(k'),
\nonumber
\\
& \Bigl\langle\delta_0(\bfp_1)\delta_0(\bfp_2)\delta_{\rm N\mbox{-}body}(\bfk')\Bigr\rangle = 
2\,(2\pi)^3\,\delta_{\rm D}(\bfp_{12}+\bfk')\,\Gamma^{(2)}(\bfp_1,\bfp_2)\,P_0(p_1)P_0(p_2),
\nonumber
\\
& \Bigl\langle\delta_0(\bfp_1)\delta_0(\bfp_2)\delta_0(\bfp_3)\delta_{\rm N\mbox{-}body}(\bfk')\Bigr\rangle = 
3!\,(2\pi)^3\,\delta_{\rm D}(\bfp_{123}+\bfk')\,\Gamma^{(3)}(\bfp_1,\bfp_2,\bfp_3)\,P_0(p_1)P_0(p_2)P_0(p_3)
\nonumber
\\
& \quad +\,(2\pi)^6\,\Bigl\{
\delta_{\rm D}(\bfp_1+\bfk')\delta_{\rm D}(\bfp_{23})\,P_0(p_2) 
+ \delta_{\rm D}(\bfp_2+\bfk')\delta_{\rm D}(\bfp_{13})\,P_0(p_1) 
+ \delta_{\rm D}(\bfp_3+\bfk')\delta_{\rm D}(\bfp_{12})\,P_0(p_3) 
\Bigr\}\Gamma^{(1)}(k')P_0(k').
\nonumber
\end{align}
The last step is to substitute these expressions into Eq.~(\ref{eq:cross_pk_LHS}). We finally obtain
\begin{align}
\PkCross^{\rm (1)}(k) &=\Gamma^{(1)}\,P_0(k),
\\
\PkCross^{\rm (2)}(k) &=2\int \frac{d^3\bfp}{(2\pi)^3}\,F_2(\bfp,\bfk-\bfp)\, \Gamma^{(2)}(\bfp,\bfk-\bfp)\,P_0(p)P_0(|\bfk-\bfp|),
\\
\PkCross^{\rm (3)}(k) &=6\int \frac{d^3\bfp d^3\bfq}{(2\pi)^6}\,F_3(\bfp,\bfq,\bfk-\bfp-\bfq)\, \Gamma^{(3)}(\bfp,\bfq,\bfk-\bfp-\bfq)\,P_0(p)P_0(q)P_0(|\bfk-\bfp-\bfq|)
\nonumber\\
&\quad+3\, \Gamma^{(1)}(k)P_0(k)\int\frac{d^3\bfp}{(2\pi)^3} \,F_3(\bfk,\bfp,-\bfp)\,P_0(p).
\end{align}

For a quantitative prediction of $r_{\rm cross}$ and $R_{\rm cross}$ based on the expressions above, we need an explicit expression for the multi-point propagators $\Gamma^{(p)}$, which are non-perturbative statistical quantities. Here, we adopt the proposition made by \cite{Bernardeau:2011dp,Taruya:2012ut}, which gives a regularized propagator that interpolates the SPT result at low-$k$ and the expected resummed behavior at high-$k$. The expressions relevant for the two-loop calculations are
\begin{align}
&\Gamma^{(1)}(k) = \Bigl\{1+\alpha_k+\frac12\alpha_k^2 + \Gamma_{\rm 1\mbox{-}loop}^{(1)}(k)(1+\alpha_k) +\,\Gamma_{\rm2\mbox{-}loop}^{(1)}(k)\Bigr\}\exp\left(-\alpha_k\right),
\label{eq:gamma1_reg}\\
&\Gamma^{(2)}(\bfk_1,\bfk_2) = \Bigl\{(1+\alpha_k)F_2(\bfk_1,\bfk_2)+ \Gamma_{\rm1\mbox{-}loop}^{(2)}(\bfk_1,\bfk_2)\Bigr\}\exp\left(-\alpha_k\right),
\label{eq:gamma2_reg}\\
&\Gamma^{(3)}(\bfk_1,\bfk_2,\bfk_3) = F_3(\bfk_1,\bfk_2,\bfk_3)\exp\left(-\alpha_k\right),
\label{eq:gamma3_reg}
\end{align}
with the quantity $\alpha_k$ given by 
\begin{align}
 \alpha_k = \frac{k^2\sigma_{\rm d}^2}{2}\,; \quad
\sigma_{\rm d}^2=\int \frac{dq}{6\pi^2}\,P_0(q).
\label{eq:alpha_k}
\end{align}
Here, the function $\Gamma_{n\mbox{-}{\rm loop}}^{(p)}$ is the $(p+1)$-propagator computed with SPT calculations at $n$-loop order (e.g., \cite{Taruya:2012ut}):
\begin{align}
&\Gamma^{(p)}_{n\mbox{-}{\rm loop}}(\bfk_1,\cdots,\bfk_p)=c^{(p)}_n
\int\frac{d^3\bfp_1\cdots d^3\bfp_n}{(2\pi)^{3n}}
 F_{(p+2n)}(\bfp_1,-\bfp_1,\cdots,\bfp_n,-\bfp_n,\bfk_1,\cdots,\bfk_p)\,P_0(p_1)\cdots P_0(p_n)
\end{align}
with the coefficient $c^{(p)}_n$ given by $_{p+2n}C_p\,(2n-1)!!$. The explicit calculations of all the expressions given above are made with public PT code, {\tt RegPT} \cite{Taruya:2012ut}.

\bibliographystyle{apsrev}


\end{document}